# Assessing the nature of nanoscale ferroelectric domain walls in lead titanate multilayers

Edoardo Zatterin[1†], Petr Ondrejkovic[2†], Louis Bastogne[3†], Céline Lichtensteiger[4], Ludovica Tovaglieri[4], Daniel A. Chaney[1], Alireza Sasani[3], Tobias Schülli[1], Alexei Bosak[1], Steven Leake[1], Pavlo Zubko[5,6], Philippe Ghosez[3], Jirka Hlinka[2], Jean-Marc Triscone[4], Marios Hadjimichael[4,7*]

1 ESRF—The European Synchrotron, 71 Avenue des Martyrs, 38000 Grenoble, France
2 Institute of Physics of the Czech Academy of Sciences, Na Slovance 2, 18221 Praha 8, Czech Republic
3 Theoretical Materials Physics, Q-MAT, Université de Liège, Allée du 6 août, 17, B-4000 Sart Tilman, Belgium
4 Department of Quantum Matter Physics, University of Geneva, 24 Quai Ernest-Ansermet, 1211 Geneva, Switzerland
5 Department of Physics and Astronomy, University College London, Gower Street, London WC1E 6BT, United Kingdom
6 London Centre for Nanotechnology, 17-19 Gordon Street, London WC1H 0AH, United Kingdom
7 Department of Physics, University of Warwick, Coventry, CV4 7AL, United Kingdom
*Corresponding author: marios.hadjimichael@warwick.ac.uk
†These authors contributed equally: Edoardo Zatterin, Petr Ondrejkovic, Louis Bastogne

*Abstract*

The observation of unexpected polarisation textures such as vortices, skyrmions and merons in various oxide heterostructures has challenged the widely accepted picture of ferroelectric domain walls as being Ising-like. Bloch components in the 180° domain walls of $PbTiO_3$ have recently been reported in $PbTiO_3/SrTiO_3$ superlattices and linked to domain wall chirality. While this opens exciting perspectives, the ubiquitous nature of this Bloch component remains to be further explored. In this work, we present a comprehensive investigation of domain walls in $PbTiO_3/SrTiO_3$ superlattices, involving a combination of first- and second-principles calculations, phase-field simulations, diffuse scattering calculations, and synchrotron based diffuse x-ray scattering. Our theoretical calculations highlight that the previously predicted Bloch polarisation in the 180° domain walls in $PbTiO_3/SrTiO_3$ superlattices might be more sensitive to the boundary conditions than initially thought and is not always expected to appear. Employing diffuse scattering calculations for larger systems we develop a method to probe the complex structure of domain walls in these superlattices via diffuse x-ray scattering measurements. Through this approach, we investigate depolarization-driven ferroelectric polarization rotation at the domain walls. Our experimental findings, consistent with our theoretical predictions for realistic domain periods, do not reveal any signatures of a Bloch component in the centres of the 180° domain walls of $PbTiO_3/SrTiO_3$ superlattices, suggesting that the precise nature of domain walls in the ultrathin $PbTiO_3$ layers is more intricate than previously thought and deserves further attention.

*Introduction*

The large strain-polarisation coupling intrinsic to ferroelectricity had historically led researchers to believe that rotations of the spontaneous polarisation away from the polar axis, which could reduce the depolarisation field energy, were forbidden [1]. In the last two decades, however, both theoretical predictions [2,3] and experimental studies using aberration-corrected transmission electron microscopy [4,5] have demonstrated that such rotations are in fact possible. This realisation motivated a search for unconventional polarisation arrangements in ferroelectric thin films and multilayers using scanning transmission electron microscopy (STEM), leading to the observation of a cornucopia of such arrangements, including ordered flux-closure domains [6,7], polar skyrmions [8], polar merons [9], and two-dimensionally ordered incommensurate structures [10].

Combined with predictions of non-Ising character at 180° domain walls [11,12], these findings have suggested the possibility of ferroelectric domain walls acting as distinct two-dimensional entities with their own phase transitions and switchable polarisation [13]. Recent work using 4D-STEM and resonant soft x-ray diffraction reported the observation of Bloch components at 180° domain walls of $PbTiO_3$ and has highlighted how this would additionally result in domain wall chirality [8], both of which are now considered ubiquitous. However, only limited experimental evidence of such polarisation components exists so far, due to the challenges in experimentally probing ferroelectric domain walls associated with their nanoscale nature. Furthermore, other works using different experimental techniques, including second harmonic generation, question this picture of chiral walls in $PbTiO_3$-based films [14,15].

In this work, we address the nature of 180° ferroelectric domain walls in ultrathin $PbTiO_3$ layers using a combination of state-of-the-art experimental and theoretical techniques. We employ first- and second-principles calculations within the generalised gradient approximation to simulate both bulk $PbTiO_3$ and $PbTiO_3/SrTiO_3$ superlattices. For bulk $PbTiO_3$, we confirm the previous predictions of Bloch components at 180° domain walls. For $PbTiO_3/SrTiO_3$ superlattices, on the other hand, we find that the Bloch polarisation is extremely sensitive to the domain period of the ferroelectric layer and could disappear completely for domain periods energetically favoured in realistic superlattices. In addition, and in contrast to predictions for bulk $PbTiO_3$, we find the Bloch polarisation in the $PbTiO_3/SrTiO_3$ system to be extremely sensitive to the magnitude of epitaxial strain imposed.

Utilising a combination of phase-field simulations, diffuse scattering calculations, and diffuse x-ray scattering measurements, we develop a method to detect and distinguish between nanoscale Bloch and depolarisation-driven, flux-closure polarisation rotations in $PbTiO_3/SrTiO_3$ superlattices. We measure diffuse x-ray scattering in such superlattices down to 2.2 K and find diffuse signal fitting our simulations for flux-closure rotations, showing that these methods can be used to deduce the nature of periodic domain walls in ferroelectric layers. However, we do not find any experimental signatures of Bloch components down to 2.2 K for any of the layer thicknesses studied, and we discuss the implications of our measurements in the context of the recent studies performed on $PbTiO_3$ films.

Our theoretical and experimental work demonstrates that the presence of the Bloch component in the domain walls of $PbTiO_3$ is not guaranteed as previously thought and is instead extremely sensitive to the boundary conditions of the system. These results should serve as a starting point to inspire more investigations into the precise structure of ferroelectric domain walls in $PbTiO_3$ thin films and will further establish diffuse x-ray scattering studies as an indispensable tool for the characterisation of complex polarisation rotations.

***Part 1: First- and second-principles study of ferroelectric domain walls in PbTiO$_3$ and PbTiO$_3$/SrTiO$_3$ superlattices within the generalised gradient approximation***

Previous first- and second-principles simulations of 180° domain walls in bulk PbTiO$_3$ and PbTiO$_3$/SrTiO$_3$ superlattices have found sizeable Bloch components in their centres, with magnitudes reaching 0.6 C/m$^2$ [13,16,17][1]. Using second-principles simulations with a model constructed from first-principles data within the local density approximation (LDA), Wojdeł and Íñiguez found that the domain wall polarisation persists beyond room temperature, with its own ferroelectric-to-paraelectric transition at approximately 340 K [13]. However, the authors highlighted the limits of using an LDA-based model to describe the domain wall transition, as this approximation over-stabilises the ferroelectric polarisation in the domain walls [13]. To verify these predictions, we repeat similar first- and second-principles calculations using the GGA-PBEsol approximation, which provides a more accurate description of the physical properties of insulating perovskites compared to LDA [18,19].

We start with 180° domain walls in bulk PbTiO$_3$. For our density functional theory (DFT) calculations, we use a $12 \times 1 \times 1$ supercell (Figure 1(a)) including domains of oppositely-oriented polarisation separated by domain walls centred on PbO planes, which appears to be the most stable configuration [20]. In this supercell, the domain polarisation is in the $z$ direction, and the domain wall Bloch polarisation is in the $y$ direction. Similarly to Ref. [13], we find that the domain wall energy is lower when we allow the domain wall to polarise (173.1 mJ/m$^2$ in the Bloch case versus 179.5 mJ/m$^2$ in the Ising case). The relaxed structure and the corresponding layer-resolved polarisation profiles along $y$ and $z$ are shown in Figure 1(b) and 1(c) respectively. Our DFT calculations are therefore consistent with previous studies and predict a sizeable Bloch component in the 180° domain walls of unstrained, freestanding PbTiO$_3$ with a polarisation $P_z = 0.93$ C/m$^2$ at the centre of the domain, and a polarisation $P_y = 0.40$ C/m$^2$ at the centre of the domain wall at 0 K.

Due to the out-of-plane polarisation along $z$, we observe a relative shift $\Delta z_{Pb} = 61$ pm of the Pb ions between up and down domains [20,21], as shown in Figure 1(b). This shift is in agreement with the value calculated by Meyer and Vanderbilt [20] and comparable to the experimentally observed shift in thin films of PbTiO$_3$ [15]. Additionally, Figure 1(b) shows that, as reported from previous first-principles studies [13,16], the domain wall polarisation arises mostly from the displacement of the Pb atoms in the $y$ direction. The shift of the Pb atoms is approximately $\Delta y_{Pb} = 20$ pm (indicated in Figure 1(b)), comparable to the observed Ti displacement in the tetragonal phase of BaTiO$_3$ [22,23] and in PbTiO$_3$ at room temperature [24].

The Pb displacement along $y$ produces some pressure to elongate the unit cell along the $b$ axis. Consistently, our relaxed supercell adopts the following lattice parameters at 0 K: $a = 3.892$ Å, $b = 3.897$ Å, $c = 4.075$ Å. This implies that, when PbTiO$_3$ is strained on cubic SrTiO$_3$ (with lattice parameter of 3.889 Å as calculated using DFT), the Bloch component might be affected. To investigate this possibility, we simulate PbTiO$_3$ strained on SrTiO$_3$ by fixing the size of the $a$ and $b$ lattice constants to 3.889 Å. Interestingly, the predicted values of the polarisation remain largely unchanged, with the polarisation inside the domain increasing slightly to $P_z$ = 0.95 C/m$^2$ and the Bloch polarisation inside the domain wall decreasing only slightly to $P_y$ = 0.37 C/m$^2$ (Figure 1(c)), and with the corresponding Pb displacement in the domain wall decreasing to $\Delta y_{Pb} = 19$ pm.

As stated above, we find from our DFT GGA-PBEsol calculations that the fully relaxed configuration at 0 K with Bloch polarization is lower in energy than without Bloch polarization by 6.4 mJ/m$^2$ = 6.5

---

[1] These estimates in C/m$^2$ were calculated considering a one-unit-cell thick domain wall centred on a PbO plane.

meV/□ (for a 12x1x1 supercell – meV/□ denotes energy per unit cell surface area of the domain wall). This energy difference is similar to that calculated by Wojdeł and Íñiguez using the same GGA-PBEsol functional (4 meV/□ using a 20x1x1 supercell), but significantly smaller than that derived from their second-principles model built on the LDA (86 meV/□ using a 20x1x1 supercell) [13]. As discussed in their work, the overestimate of the energy difference by their model indicates that the real ferroelectric-to-paraelectric transition temperature in the domain wall could be significantly smaller than the predicted temperature of 340 K [13]. In order to better estimate the temperature at which the Bloch polarisation should disappear, we construct a new second-principles model relying on our GGA-PBEsol calculations (see Methods).

To compare our new second-principles model with previous DFT calculations, we first employ the same $12 \times 1 \times 1$ supercell, obtaining consistent results: the appearance of the non-zero domain wall Bloch polarisation lowers the energy by 9.3 meV/□ and gives rise to a structure where the polarization in the middle of the domain is $P_z = 0.84$ C/m², the Bloch polarisation in the domain wall is $P_y = 0.42$ C/m² and the Pb displacement in the domain wall is $\Delta y_{Pb} = 19.5$ pm. To properly estimate the temperature evolution of the Bloch component of the polarisation, we utilise a larger $20 \times 12 \times 12$ supercell (strained on SrTiO₃). Figure 1(d) shows the calculated domain wall Bloch polarisation $P_y$ and the corresponding magnitude of $\Delta y_{Pb}$ as a function of temperature for bulk PbTiO₃. We find that $P_y$ becomes zero at approximately 150 K, which, as expected, is below the temperature predicted in Refs [13] and [8] but remains quite high. Interestingly, we note that for freestanding PbTiO₃ (no imposed strain), we find a transition temperature equal to approximately 200 K.

We now move from pure PbTiO₃ to (001)-oriented PbTiO₃/SrTiO₃ superlattices (from now on we will denote these systems with (PTO$_n$|STO$_m$)$_N$, where n and m are the number of unit cells of PbTiO₃ and SrTiO₃ per period respectively, and N is the number of repetitions). The PbTiO₃ layer is now sandwiched between SrTiO₃ layers, limiting the extension of the domain wall in the $z$ direction. In this case, the incomplete screening of the depolarizing field by the SrTiO₃ layer typically gives rise to flux-closure domains [21]. Using the second-principles model of Wojdeł and Íñiguez constructed on LDA data, Ortiz *et al.* explored the evolution of the polar textures with layer thicknesses, epitaxial strain and temperature [17,25]. They predicted the presence of a Bloch component of polarization in domain walls up to room temperature, which further appeared as a key feature to explain the formation of polar skyrmions [8]. However, other simulations of PbTiO₃ thin films using LDA [26], and alternative models for domain walls in (PTO$_{10}$|STO$_{10}$)$_\infty$ relying on the GGA-PBEsol functional [27] did not report any Bloch component. To gain more insight, we take our PbTiO₃ and SrTiO₃ second-principles bulk models built on GGA-PBEsol data and combine them according to previously proposed methodology [28] to reinvestigate PbTiO₃/SrTiO₃ superlattices.

As a prototypical example, we consider (PTO$_6$|STO$_6$)$_\infty$ superlattices epitaxially grown on a SrTiO₃ cubic substrate. For such layer thicknesses, our calculations find that the equilibrium domain period ($\Lambda_d$) is approximately 18 unit cells (see Supplementary Figure 1 - for comparison, other calculations found an equilibrium domain period of 12-14 unit cells at these thicknesses [25]). Importantly, as shown in Figure 2, while we observe the presence of a Bloch component in the domain wall for small lateral domain sizes ($P_y = 0.27$ C/m² for $\Lambda_d = 8$ unit cells), this component progressively reduces with increasing $\Lambda_d$ and disappears completely for domain periods larger than 14 unit cells. Consistently, we find that the Bloch component disappears completely at periods lower than the equilibrium domain periods of superlattices with different layer thicknesses. Figure 2(c) shows a plot of the magnitude of the Bloch component as a function of domain period for bulk PbTiO₃ and for superlattices with varying thickness of PbTiO₃ and SrTiO₃, specifically (PTO$_6$|STO$_6$)$_\infty$, (PTO$_{10}$|STO$_6$)$_\infty$, and (PTO$_{10}$|STO$_{10}$)$_\infty$. At the

equilibrium domain period for each system (18, 26 and 24 unit cells respectively, marked by the vertical arrows) the magnitude of the Bloch component is zero.

The absence of Bloch polarization at these domain walls is consistent with simulations on PbTiO$_3$/SrTiO$_3$ using machine learning potentials [27]. However, it is at odds with our predictions for bulk PbTiO$_3$, for which the Bloch component is largely independent of the lateral size of the domains. The disappearance of the Bloch component in our superlattices is most likely related to the finite size of the PbTiO$_3$ layer in the simulated PbTiO$_3$/SrTiO$_3$ superlattices (in contrast to bulk PbTiO$_3$) and might originate from the interaction between the Bloch and flux-closure components of the polarization.

Importantly, we also observe that the magnitude of the Bloch component is strongly sensitive to the epitaxial strain as shown in Figure 2(d). For a domain period of 12 unit cells, the Bloch component disappears when we increase the compressive strain from 0 to -0.2% (see Methods for the definition of the lattice parameters for the PbTiO$_3$/SrTiO$_3$ system). From a modelling perspective, these observations mean that making reliable predictions remains challenging since the results might be very sensitive to the approximations that are used and the way the strain is treated in the simulations. Experimentally, it means that partial relaxation of the superlattice and local fluctuation of the strain (e.g. during preparation of lamellae for STEM analysis) might strongly affect the observations.

In summary, our calculations based on the GGA-PBEsol functional are consistent with previous studies and confirm the presence of a Bloch component in bulk PbTiO$_3$ 180° domain walls. This method allowed us to obtain a more reliable estimate of the temperature at which the Bloch component should disappear. Furthermore, our theoretical analysis supports the possibility of a Bloch component developing in the 180° domain walls of PbTiO$_3$/SrTiO$_3$ superlattices. We find a strong and previously unreported sensitivity of this component to the ferroelectric domain periodicity and the epitaxial strain. These findings suggest that the presence of a Bloch component in PbTiO$_3$/SrTiO$_3$ superlattices might not be ubiquitous and, if present, may be more challenging to detect than previously thought.

***Part 2: Phase-field simulations of labyrinthine 180° domain walls in PbTiO$_3$/SrTiO$_3$ superlattices***

To explore more realistic domain structures with labyrinthine domain wall orientations, a larger volume of the material must be calculated. One of the most convenient models by which polarization distributions over several millions of unit cells can be simulated is the Ginsburg-Landau-Devonshire model with long-range electrostatic interactions, a method known as phase field simulations.

Our phase-field simulations were performed using the program Ferrodo (see Methods). The system chosen is a (PTO$_{13}$|STO$_5$)$_{12}$ superlattice under a compressive epitaxial strain of $-1.5\%$ with respect to the cubic paraelectric phase of PbTiO$_3$, thus simulating the constraint imposed by a SrTiO$_3$ substrate. Figure 3(a) shows the calculated three-dimensional structure, which exhibits a labyrinthine domain pattern with 180° domain walls and a domain period equal to approximately 27 unit cells. This type of labyrinthine structure is consistent with experimentally observed domain wall patterns in PbTiO$_3$ thin films and multilayers deposited on low-miscut SrTiO$_3$ substrates [8,29–33], as well as other reports of systems with modulated phases [34]. The domain walls in the simulated PbTiO$_3$/SrTiO$_3$ system are mostly aligned along the whole thickness of the superlattice, signifying a large degree of coherence in the out-of-plane direction, most likely as a result of structural and/or electrostatic coupling through the SrTiO$_3$ layers [35].

A cross-section in the $x$-$z$ plane of a single simulated PbTiO$_3$ layer is shown in Figure 3(b). A flux-closure pattern is visible, characterised by a polarisation that is directed out of plane at the centres of the domains, with maximum magnitude equal to 0.7 C/m$^2$ that continuously rotates to in plane at the domain walls and top/bottom interfaces with SrTiO$_3$, with maximum magnitude equal to 0.45 C/m$^2$.

These values imply that while the term 'flux-closure' is used for simplicity, the flux does not close completely, and an excess out-of-plane polarisation remains in the centre of each domain at the interfaces between PbTiO$_3$ and SrTiO$_3$. The domains are symmetric such that the $-P_x$ and $+P_x$ regions occupy the same volume and have the same magnitudes of polarisation, in contrast to tensile-strained PbTiO$_3$ thin films and multilayers on GdScO$_3$ [6] and DyScO$_3$ [7].

The $x$-$y$ cross-sections of the same simulated PbTiO$_3$ layer at $z$ positions corresponding to the middle of the layer and the top interface with SrTiO$_3$ are plotted in the left and right panels of Figure 3(c), respectively. No in-plane polarisation components are visible in the middle of the PbTiO$_3$ layer, while at the top interface the in-plane polarisation is perpendicular to the domain walls, consistent with the flux-closure configuration inferred from Figure 3(b). Therefore, no Bloch components emerge from our phase-field simulations, in agreement with our second-principles calculations on PbTiO$_3$/SrTiO$_3$ superlattices with their equilibrium domain periods.

For the following discussion it is useful to consider an artificially designed domain structure identical to the above simulated domain structure, but with Ising domain walls replaced by Bloch ones. We have constructed this artificial structure by using the as-received polarisation pattern and superposing onto it an additional in-plane polarisation field, calculated from the gradient of the as-received out-of-plane polarisation according to the expressions: $P'_x = P_x - \frac{\partial P_z}{\partial y}|dy|$, and $P'_y = P_y + \frac{\partial P_z}{\partial x}|dx|$. The artificial Bloch polarisation obtained is approximately equal to 0.55 C/m², comparable to our first-principles predictions for a domain wall in bulk PbTiO$_3$. The spatial distribution of polarisation within this homochiral Bloch structure is plotted in Figure 3(d). Here, the Bloch polarisation parallel to the periodic domain walls is clearly visible in the cut through the middle of the PbTiO$_3$ layer (left panel).

### *Part 3: Calculation of diffuse scattering from phase-field polarisation patterns*

Phase-field simulation data can be used to model the characteristic diffuse scattering fingerprints from different polarization patterns. The scattered x-ray amplitude $F(\mathbf{Q})$ from a ferroelectric crystal with 180° domain walls can be computed as:

$$F(\mathbf{Q}) = \sum_{j}^{crystal} f_j \exp[-i\mathbf{Q} \cdot (\mathbf{r}_{0j} + \delta\mathbf{r}_j)], \quad (1)$$

where the sum is over all atoms in the crystal structure, $\mathbf{Q}$ is the scattering vector, $f_j$ is the form factor of atom $j$ in the parent centrosymmetric structure, and $\delta\mathbf{r}_j$ the displacement of the atom from its equilibrium position $\mathbf{r}_{0j}$ with respect to the centrosymmetric state due to the local polar distortions $\mathbf{p}_a$ and local acoustic displacements $\mathbf{u}$ (see Methods).

This equation entails that the scattered intensity will be modulated due to the presence of 180° domain walls when the scalar product between the scattering vector $\mathbf{Q}$ and the atomic displacement $\delta\mathbf{r}_j$ is non-zero [29,30]. Therefore, for an ideal labyrinthine 180° domain pattern with no in-plane polarisation components we expect no diffuse signal around $HK0$ pseudocubic Bragg peaks for which $Q_z = 0$, and an isotropic ring of diffuse intensity around $HKL$ reflections with $L \neq 0$, for which $Q_z \neq 0$ [33]. Conversely, when in-plane components are present and $\delta\mathbf{r}_j$ has non-zero $x$ and $y$ components, diffuse signal will also be observed around $HK0$ reflections and will necessarily be anisotropic because of the scalar product in Equation 1. The exact shape of this anisotropy will depend on whether the polarisation components are perpendicular or parallel to the domain walls (Supplementary Figure 2), and therefore encodes information on the precise nature of the polarisation texture being probed.

We thus proceed to calculate the diffracted intensity $|F(\mathbf{Q})|^2$ from our simulated Ising-like and artificially induced Bloch wall domain patterns by substituting the corresponding atomic positions into

Equation 1 (see Methods). The resulting intensity distribution in the $0KL$ reciprocal space plane arising from the Ising-like structure is shown in Figure 4(a). Along the out-of-plane direction, the repeating unit of this superlattice consists of 18 unit cells, giving rise to 18 superlattice Bragg peaks up to $L = 1$. For clarity throughout the manuscript, we index superlattice reflections with lowercase indices $hkl$, whereas for the SrTiO$_3$ substrate peaks in the subsequent experimental sections we use $HKL$. As expected, superlattice Bragg peaks with $Q_z \neq 0$ are accompanied by diffuse signal due to the periodicity of the 180° domain walls in our simulated labyrinthine structure in Figure 3(a). The distance between the diffuse and the superlattice peaks is $\Delta H = 0.037$ r.l.u., corresponding to a simulated domain period $\Lambda_d = \frac{1}{\Delta H} = 27$ u.c. (where u.c. denotes the unit cell of the phase-field derived structure with the reference lattice constant 0.4 nm, see Methods).

Figure 4(b) and Figure 4(c) show the $hk0$ (top) and $hk1$ (bottom) planes obtained from simulated patterns with the Ising-like flux-closure and artificially induced Bloch wall textures, respectively. The $hk1$ planes of both textures are qualitatively similar and are consistent with the scenario sketched in Supplementary Figure 2(b), which depicts the diffuse signatures of a domain pattern with polarisations perpendicular to the plane of the domain walls. This diffuse signal comes from in-plane components of the flux-closure pattern near the PbTiO$_3$/SrTiO$_3$ interfaces (see Figure 3). On the other hand, inspection of the $hk0$ planes reveals that the flux-closure pattern gives rise to weak second order diffuse intensity with maxima along the scattering vector $\boldsymbol{Q}$, whereas the pattern including Bloch polarisation gives rise to additional strong transverse diffuse signal. This strong transverse diffuse signal is consistent with a domain pattern with polarisations parallel to the plane of the domain walls (i.e. Bloch walls) as depicted in Supplementary Figure 2(a).

Figures 4(d) and 4(e) are detailed plots of the calculated diffuse signal in the $HK$ plane around the 100 ($L = 0$), 101 ($L = 0.06$ r.l.u.) and 1 0 54 ($L = 3$) superlattice peaks for the cases of flux-closure domain walls with no polarisation in the centre of the domain wall and for flux-closure walls with Bloch polarisation in the centre of the domain wall respectively. The middle and bottom rows are calculations in which contributions from local polar distortions $\boldsymbol{p_a}$ (indicated as polarisation only) and local acoustic displacements $\boldsymbol{u}$ (indicated as strain only) have been isolated.

The separated contributions give similar diffuse features. The maps derived from local acoustic displacements ("strain only") show broadening of Bragg and superlattice reflections (which will be discussed further below), and weak second order diffuse intensity. The second order diffuse scattering, corresponding to a period equal to $\frac{\Lambda_d}{2} = 13 \, u.c.$, arises from the electrostriction related strain contrast, which gives the mean size of simulated domains. However, when we consider only the effect of polar distortions ("polarisation only"), we see that the diffuse signal around the 100 superlattice peak disappears, as the contributions to the structure factor from the top and bottom part of the PbTiO$_3$ layer cancel (see variation of the $P_x$ component in Figure 3(b)).

Our calculations demonstrate that by measuring diffuse signal around Bragg peaks with $L = 0$ we can distinguish between an Ising configuration or a configuration with Bloch polarisation in the 180° domain walls of PbTiO$_3$ layers in PbTiO$_3$/SrTiO$_3$ superlattices. We note that similar guidelines were successfully applied to detect Bloch (and Néel) domain walls in magnetic single crystals and multilayers using resonant soft x-ray diffraction (RSXD) [36–40]. Furthermore, we also show that diffuse signal around peaks with non-zero but small values of $l$ is sensitive to the nanoscale polarisation rotation driven by the depolarising field in the 180° domain walls of ultrathin PbTiO$_3$. Careful study of the diffuse scattering around these peaks will therefore allow for such nanoscale polarisation rotation to be probed via x-ray scattering techniques.

*Part 4: Diffuse x-ray scattering studies of the structure of ferroelectric domain walls in PbTiO₃*

To determine the polarisation configuration of the ferroelectric domain walls in PbTiO$_3$/SrTiO$_3$ superlattices deposited on SrTiO$_3$ substrates we perform measurements at the ID28 diffuse x-ray scattering end-station at ESRF (see Methods) and compare the results to the schematic outlines above. The $0KL$ reciprocal space plane of a (PTO$_{16}$|STO$_5$)$_{11}$ superlattice at room temperature is shown in Figure 5(a). The coordinates are expressed in reciprocal lattice units (r.l.u.) of SrTiO$_3$, where $c = 3.905$Å [2]. Note that unlike fine period superlattices, only peaks at integer values of $K$ are present, indicating that there is no doubling of the pseudocubic unit cell along the in-plane directions due to antiferrodistortive structural ordering and/or improper ferroelectricity [41,42].

Figures 5(b) and 5(c) show the superlattice peaks around $K = -1$ and $K = 1$ in more detail. The repeating superlattice unit consists of 21 perovskite unit cells with an average out-of-plane lattice parameter $c_{SL} = \frac{c_{STO} n_{STO} + c_{PTO} n_{PTO}}{n_{STO} + n_{PTO}}$, where $c_{STO}$, $c_{PTO}$ are the out-of-plane lattice parameters of the SrTiO$_3$ and PbTiO$_3$ layers respectively, and $n_{STO}$, $n_{PTO}$ are the numbers of unit cells of SrTiO$_3$ and PbTiO$_3$ per repetition. Due to the similar values of $c_{STO}$ and $c_{PTO}$, there are 21 superlattice Bragg peaks up to $L = 1$, and these peaks are indexed accordingly. The first superlattice Bragg ($l = 1$) then occurs at $L = 0.046$ r.l.u., and the 21$^{st}$ superlattice Bragg peak ($l = 21$) at $L = 0.970$ r.l.u. The diffuse signal arising from periodic 180° domains occur at $\Delta K = 0.035$ r.l.u., and correspond to a domain period $\Lambda_d = 110$ Å.

Figure 5(d) shows a subsection of the hk0 plane for the (PTO$_{16}$|STO$_5$)$_{11}$ superlattice ($l = 0$, $L = 0$). No diffuse signal is observed around the $100$, $0\bar{1}0$ and $1\bar{1}0$ peaks of the superlattice, in line with other x-ray diffraction experiments on ferroelectric domains in PbTiO$_3$ thin films deposited on SrTiO$_3$ [29]. Conversely, around the $hk1$ superlattice peaks ($l = 1$, $L = 0.046$ r.l.u.), shown in Figure 5(e), we observe clear diffuse intensity arising from the periodic domain structure, with minima in intensity along a direction perpendicular to the in-plane component of the scattering vector $Q_{IP}$. Given that the second-order signal around the $hk0$ peaks in Figure 4(d) is predicted to be very weak, and is likely not detectable experimentally, we find that the experimentally observed diffuse patterns are entirely consistent with the nanoscale flux-closure structure found as the ground state of our phase-field simulations.

Additional higher resolution measurements performed at beamline ID01, ESRF on a (PTO$_{13}$|STO$_5$)$_{12}$ superlattice (Figure 6) confirm the absence of diffuse scattering around the $l = 0$ reflections and identical anisotropic diffuse scattering around the $l = 1$ peaks. Furthermore, with the improved resolution we can observe additional structured diffuse signal close to the Bragg peaks with characteristic butterfly or rod shapes. This signal is consistent with Huang-like scattering [43] observed in a variety of systems; from materials with point defects and defect clusters [44], to systems exhibiting Jahn-Teller distortions [45]. Such scattering is not produced directly from local displacements, but from the long-range strain field that arises in the crystal lattice in response. Diffuse signal (broadening) with similar characteristic shapes is well captured by the acoustic displacement "strain" term of our calculations (Figure 4) and most likely arises from the long-range response of the crystal lattice to the Pb and Ti displacements. However, the precise origin and details of this diffuse signal are beyond the scope of this work and therefore, other than to note the general good agreement between theory and experiment, no further discussion is provided.

---

[2] To convert from r.l.u. to Å$^{-1}$, the values should be multiplied by $\frac{2\pi}{3.905}$ Å$^{-1}$.

With increasing $l$, the diffuse scattering from the periodic domain structure becomes more isotropic. Figure 5(f) shows the $hk21$ plane ($l = 21$, $L = 0.970$ r.l.u.), where a diffuse ring is present, similar to previous reports on PbTiO$_3$ thin films [29,30] and PbTiO$_3$/SrTiO$_3$ superlattices deposited on SrTiO$_3$ substrates [8,33,46]. However, the intensity is not entirely isotropic, with enhanced intensity along the $\langle 100 \rangle$ directions, indicating a preference for $\{100\}$-oriented domain walls. Additionally, the four-fold symmetry of the diffuse signal is further reduced as the diffuse signals with positive and negative $\Delta K$ have different intensities, shown clearly in Supplementary Figure 3. Further diffuse scattering calculations (not shown here) demonstrate that this asymmetry is consistent with the combination of periodic out-of-plane and in-plane atomic displacements due to the in-plane polarisation rotation at the 180° domain walls of the PbTiO$_3$ layers. Note that the apparent anisotropy around the $1\bar{1}21$ peak is an artifact arising from integration over the detector gap. Almost fully isotropic diffuse signal around this reflection was confirmed by measurements at a different detector position.

High-temperature measurements of the $0\bar{1}1$ and $0036$ ($l = 36$, $L = 1.944$ r.l.u.) superlattice peaks for the (PTO$_{13}$|STO$_5$)$_{12}$ system (Supplementary Figure 5) show that the diffuse signal around both disappears at 723 K, due to the disappearance of the ferroelectric polarisation and the 180° domain structure. Reducing the temperature to 2.2 K does not lead to any striking changes in the diffuse signal around the $0\bar{1}1$ superlattice peak, nor the formation of diffuse signal around the $0\bar{1}0$ superlattice peak (Supplementary Figure 6). However, subtle changes occur in the positions of the diffuse signal maxima around the $0\bar{1}1$ peak, with a gradual change in their position starting below 100 K. These changes are summarised in the Supplementary Materials (see Supplementary Figures 5 and 6 with associated discussion).

Our experimental observations indicate that for all temperatures, the absence of diffuse scattering at $L = 0$ rules out a structure with periodic Bloch components in our PbTiO$_3$/SrTiO$_3$ superlattices, consistent with both second principles and phase-field calculations for systems with domain periods larger than 14 unit cells. However, our results still leave open the possibility of a disordered Bloch structure with a random direction of the Bloch polarisation, due to a lack of macroscopic coherence of the Bloch component in three dimensions. Naively, this would agree with the simulations in PbTiO$_3$ and two-dimensional PbTiO$_3$/SrTiO$_3$ supercells when a Bloch component is present, which show that there is no significant difference between configurations with parallel and antiparallel Bloch polarisations in adjacent walls (percentage difference in energy of $3 \times 10^{-7}$ %). Nevertheless, we note that such a small difference in energy does not necessarily exclude all macroscopic correlations: labyrinthine domain structures similar to the one in Figure 3(a) can be created using one long meandering domain wall, and the macroscopic coherence of the Bloch component in such cases will not depend on the direct coupling between adjacent walls, but on the coherence of the Bloch component along the wall itself. Completely disordered Bloch polarisation is also not consistent with recent observations of macroscopic chirality in 180° domain walls interpreted as arising from macroscopically ordered Bloch components in PbTiO$_3$/SrTiO$_3$ superlattices [8,47,48]. The reason for the discrepancy between our investigations and such experimental works is still uncertain. The sensitivity of the Bloch component to local strain and domain period, as shown in our second principles calculations, could be crucial for the behaviour of the polarisation inside the domain walls. Additionally, different sample fabrication techniques (the aforementioned works have studied samples fabricated using pulsed laser deposition, whereas our work focuses on samples grown using off-axis sputtering), as well as the precise nature of the heterostructure (e.g. the use of a bottom SrRuO$_3$ layer) could play an important role. The reason for this discrepancy between our study and previous works therefore deserves more attention and should be investigated further in future studies.

Finally, our x-ray diffraction measurements indicate that for small, non-zero values of $Q_z$ we detect an in-plane modulation of the atomic displacements occurring in a perpendicular direction to the 180° domain wall plane. The results are consistent with calculated diffuse scattering in the $hk1$ plane (Figures 4(b) and (d)). This diffuse scattering is a signature of the rotation of the atomic displacements at the 180° domain walls of PbTiO$_3$ driven by the depolarising field. Our measurements therefore show that signatures of such atomic displacements can be observed in diffuse x-ray scattering, providing a macroscopic technique that can be used to study the rotation of spontaneous polarisation.

*Conclusions*

To conclude, this work focused on understanding the structure of ferroelectric domain walls in PbTiO$_3$ using a combination of first- and second-principles calculations, phase-field simulations, diffuse scattering calculations, and synchrotron based diffuse x-ray scattering measurements. Our first- and second-principles calculations confirmed that a Bloch domain wall configuration is expected in bulk PbTiO$_3$ when strained on SrTiO$_3$, with a sizeable polarisation and a characteristic transition temperature of approximately 150 K. The structure of this Bloch wall is characterised by correlated displacements of the Pb cations in the direction of the Bloch polarisation. On the other hand, second-principles calculations showed that the Bloch polarisation in PbTiO$_3$/SrTiO$_3$ superlattices is extremely sensitive to the domain period and the imposed epitaxial strain and could disappear at their equilibrium domain periods.

Using the structural signature of the Bloch polarisation, we calculated diffuse x-ray scattering from phase field patterns with periodic non-Ising domain walls, with our phase-field simulations suggesting that the characteristic diffuse scattering patterns around peaks with zero out-of-plane ($Q_z$) components of the scattering vector would allow us to distinguish between different types of domain walls. We performed diffuse x-ray scattering measurements, and by measuring diffuse signal around low $Q_z$ peaks in PbTiO$_3$/SrTiO$_3$ superlattices, we found anisotropic diffuse intensity. Diffuse scattering fingerprints of phase-field simulated patterns showed that this anisotropic diffuse intensity arises due to the ferroelectric polarisation configuration associated with the flux-closure structure of the 180° domain walls, characterised by a rotation of polarisation, a result of the inhomogeneous depolarising fields in the PbTiO$_3$ layers. We observed no signature of Bloch components down to 2.2 K, consistently with our second-principles calculations.

Although our measurements do not exclude the presence of the Bloch polarisation at the domain walls of all PbTiO$_3$ films and multilayers, they emphasise the importance of the boundary conditions imposed on these systems, as well as the effect of epitaxial strain and layer thickness. The simulations and measurements in this work therefore provide new insight into the structure of ferroelectric domain walls in PbTiO$_3$-based multilayers, showing that the nature of these nanoscale objects is intricately linked to the precise structure of the studied films and multilayers. Our results emphasising this sensitivity of the Bloch component hopefully reconcile contrasting observations of the nature of the domain walls of PbTiO$_3$-based thin films and multilayers found in literature. Finally, we expect that our general guidelines for distinguishing between different types of ordered polarisation configurations in ferroelectric multilayers using diffuse x-ray scattering will be useful for determining the nature of domain walls in other systems and hope they will inspire future studies.

*Figures*

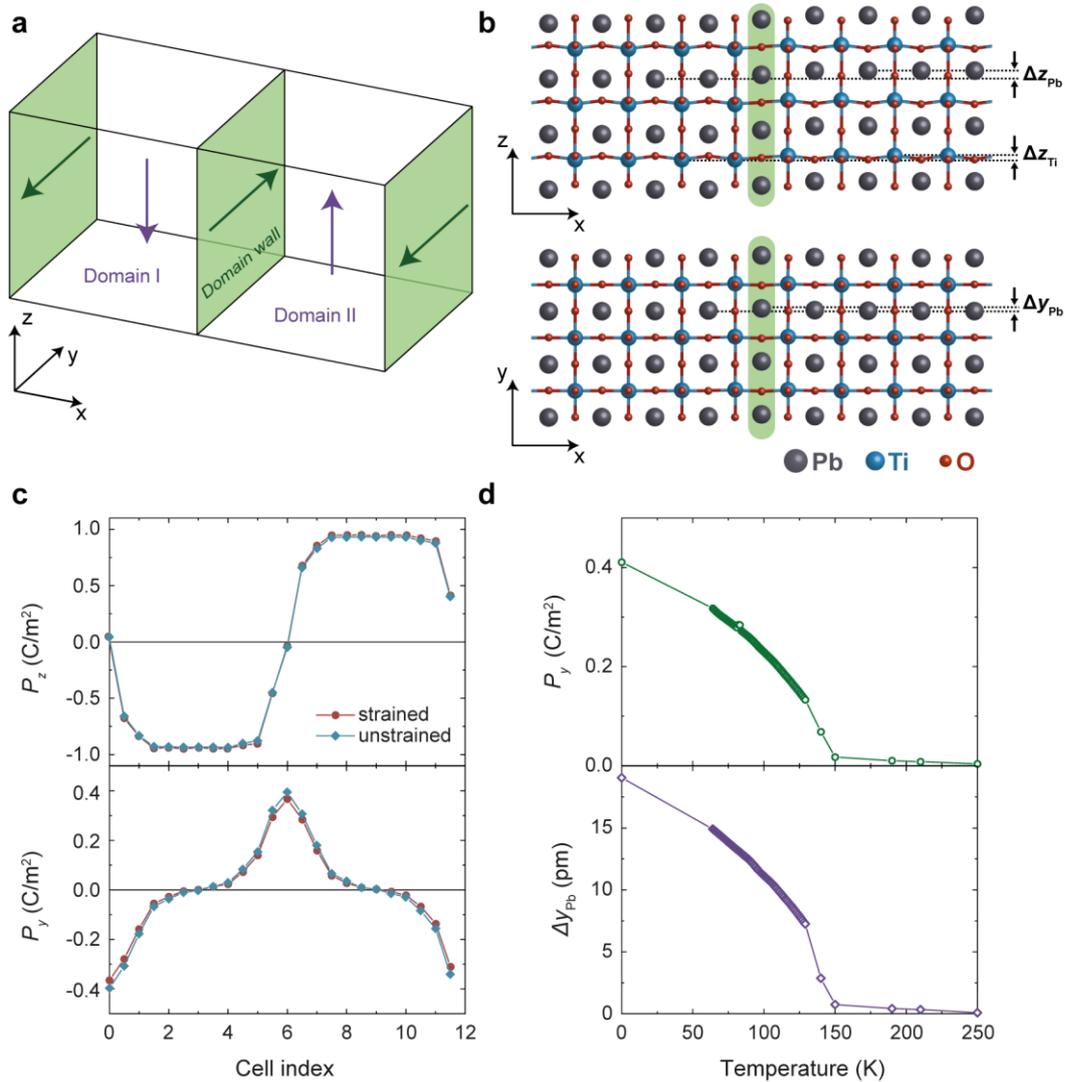

Figure 1. First- and second-principles calculations of the structure of 180° domain walls in bulk PbTiO$_3$. (a) Sketch of the relative orientations of the polarisation in the domains and domain walls of the supercells used in the first- and second-principles simulations. The polarisation in the centre of each domain points along the $z$ direction, whereas a polarisation in the plane of the wall (Bloch polarisation) points along the $y$ direction. (b) Lowest-energy structure for freestanding PbTiO$_3$, viewed in the $x-z$ (top) and $x-y$ (bottom) planes. The polarisation inside the domains leads to a displacement of the Pb and Ti cations along $z$, equal to $\Delta z_{Pb} = 61$ pm and $\Delta z_{Ti} = 32$ pm between up and down domains, respectively. The polarisation inside the domain wall leads to a Pb displacement along $y$, equal to $\Delta y_{Pb} = 20$ pm. The shaded green region marks the position of the 180° domain wall. (c) Layer-resolved polarisation in the $z$ (top) and $y$ (bottom) directions for freestanding PbTiO$_3$ (blue diamonds) and for PbTiO$_3$ strained on SrTiO$_3$ (red circles). The plot shows that the polarisation along the $y$ direction reaches values of 0.4 C/m$^2$ inside the centre of the domain walls and remains largely unchanged upon application of strain. (d) Second-principles calculations of the temperature dependence of the magnitude of the individual domain wall polarisation, $P_y$ (top), and the magnitude of the average Pb displacement in the domain wall, $\Delta y_{Pb}$ (bottom) for unstrained PbTiO$_3$, showing a ferroelectric-to-paraelectric transition occurring inside the domain wall at approximately 150 K. For the first-principles calculations, a $12 \times 1 \times 1$ supercell was used. For the second-principles calculations, a $20 \times 12 \times 12$

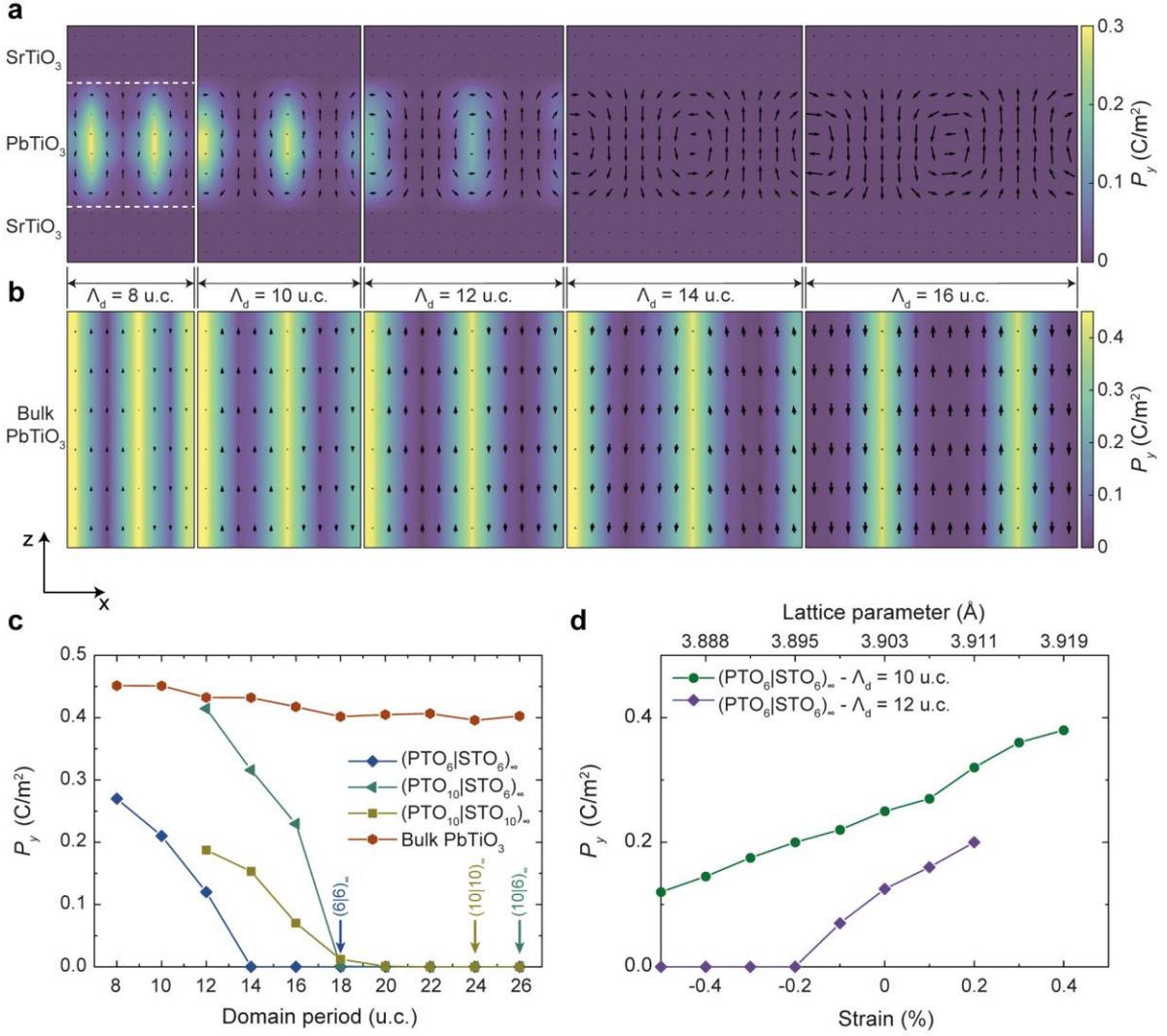

Figure 2. Second-principles calculations of the structure of 180° domain walls in PbTiO$_3$/SrTiO$_3$ superlattices. (a) Cross-sections of a (PTO$_6$|STO$_6$)$_\infty$ superlattice period viewed in the $x-z$ plane for domain periods ranging from 8 to 16 unit cells. The colour scales depict the magnitude of $P_y$ whereas the arrows mark the orientation of the spontaneous polarisation. The Bloch polarisation in the centres of the 180° domain walls decreases in magnitude with increasing domain period and disappears between 12 and 14 unit cells. (b) Domain period dependence of the Bloch polarisation in bulk PbTiO$_3$ for comparison. No significant changes in the Bloch component occur. Note the different colour scale compared to panel (a). (c) Plot of the magnitude of the Bloch polarisation in the domain walls of bulk PbTiO$_3$ (red circles), and for PbTiO$_3$/SrTiO$_3$ superlattices with different layer thicknesses as a function of the domain period. The arrows indicate the equilibrium domain periods for the PbTiO$_3$ layer in each superlattice. The Bloch polarisation is systematically zero at the equilibrium domain period. (d) Plot of the Bloch polarisation in the domain walls of the PbTiO$_3$/SrTiO$_3$ superlattice as a function of strain, for a domain period of 10 (green circles) and 12 (purple diamonds) unit cells.

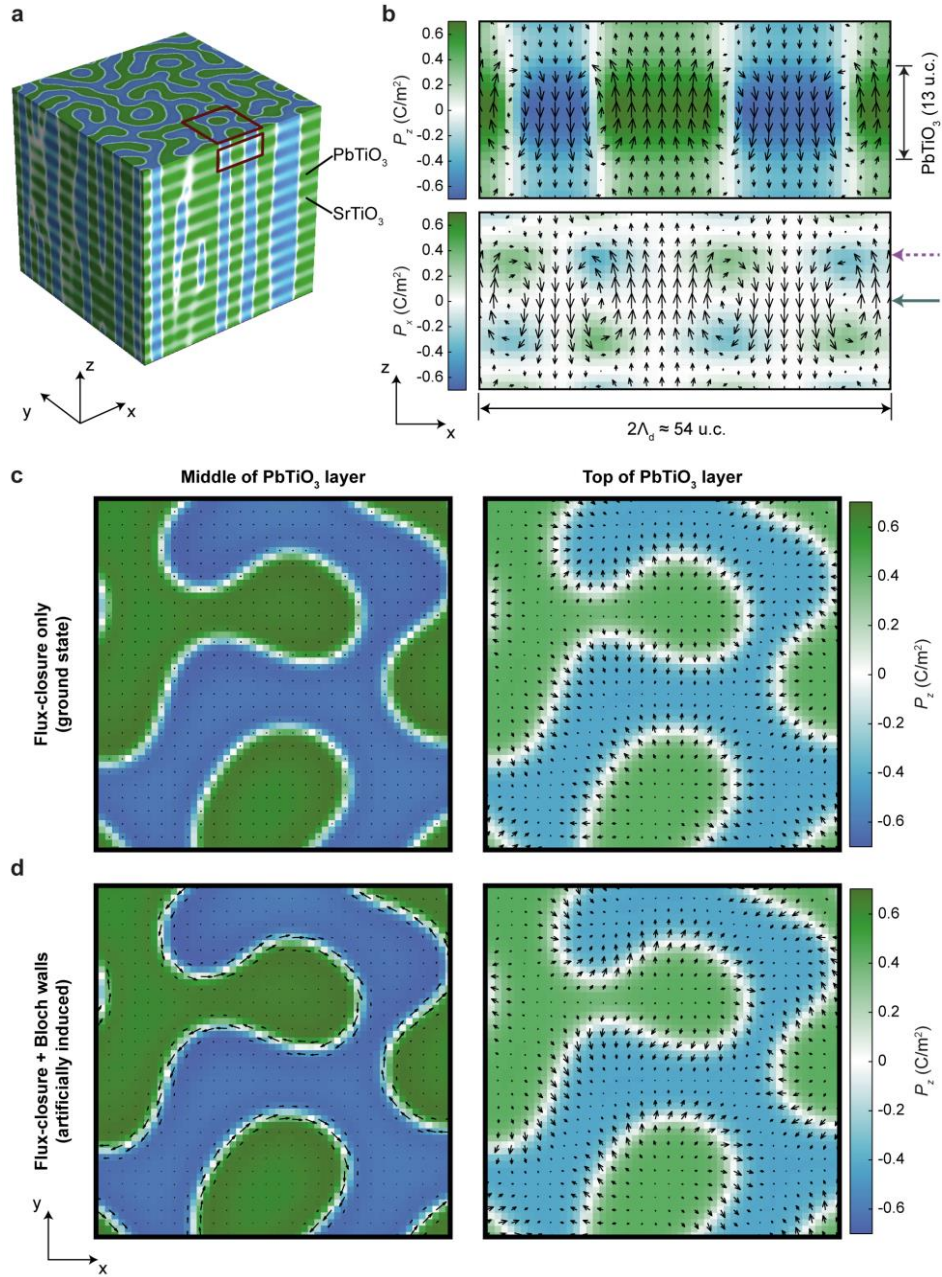

Figure 3. Phase field simulations of the domain structure in a (PTO$_{13}$|STO$_5$)$_{12}$ superlattice. (a) Simulated 3D distribution of the polarisation component in the $z$ direction, $P_z$, for a superlattice with 12 periods of 13 unit cells of PbTiO$_3$ and 5 unit cells of SrTiO$_3$, with the middle of the PbTiO$_3$ layer at the top of the supercell. (b) Cross-section of a superlattice period viewed in the $x-z$ plane in the region marked by the rectangle in panel (a). The colour scales depict the magnitude of $P_z$ (top panel) and $P_x$ (bottom panel), whereas the arrows mark the orientation of the spontaneous polarisation. The bottom panels correspond to the polarisation distribution in the $x-y$ plane for (c) the ground state structure with only a flux-closure type polarisation rotation, and (d) a structure with artificially induced Bloch components in the centres of the domain walls. The left and right hand sides of each panel are cuts through the middle and top of the PbTiO$_3$ layer, marked by the solid and dashed arrows in panel (b), respectively. The colour scales correspond to the magnitude of $P_z$, and the arrows to the direction of the polarisation.

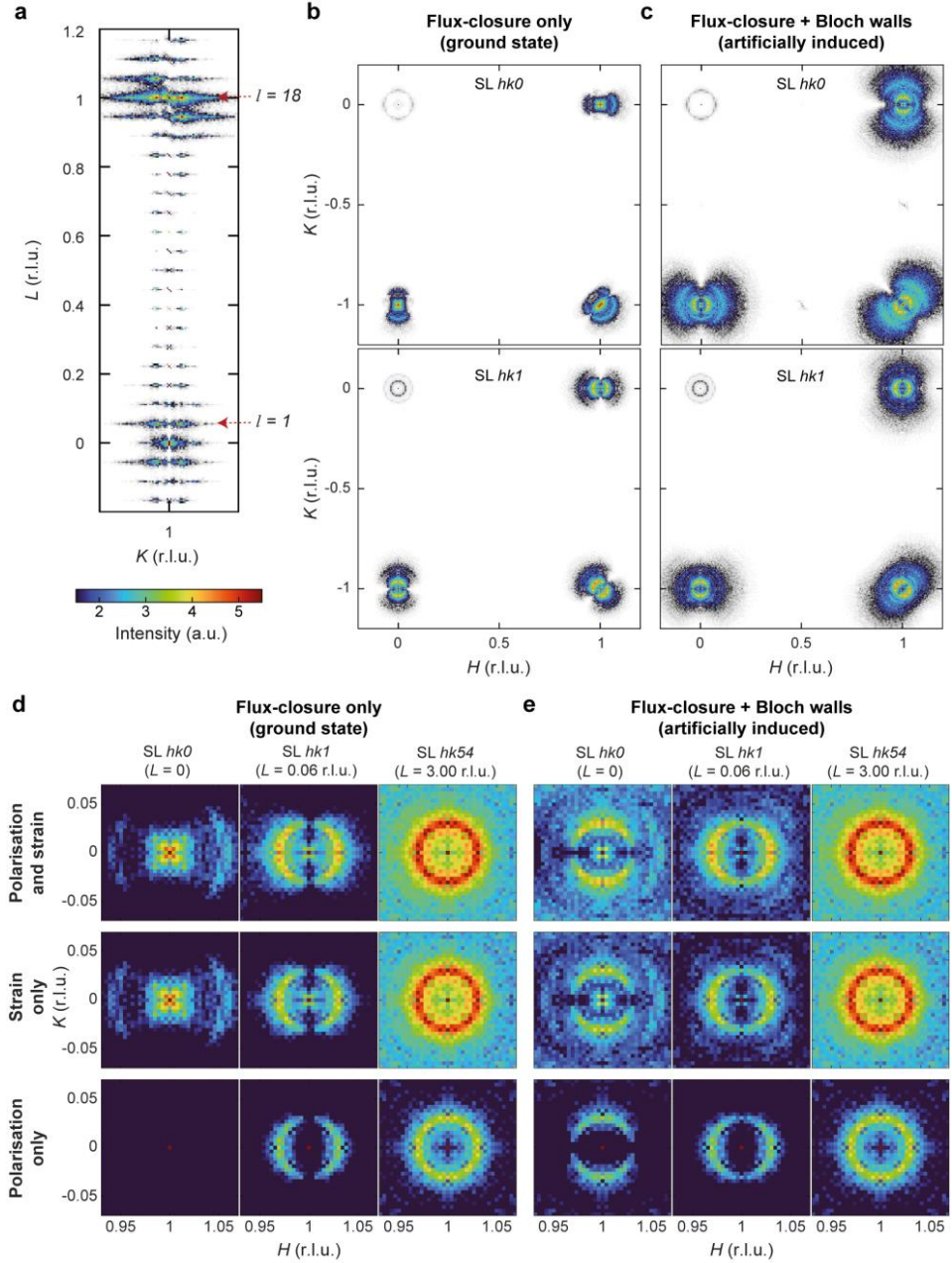

Figure 4. Diffuse scattering calculations for nanoscale flux-closure domain walls for the Ising and artificially induced Bloch case, and the comparison of lattice distortion versus polarisation. Calculated diffuse scattering intensity map in the $0KL$ plane of the $(PTO_{13}|STO_5)_{12}$ superlattice simulated in Figure 3. $l = 1 \ldots 18$ superlattice peaks occur between $L = 0$ and $L = 1$. Diffuse scattering appears around each superlattice Bragg peak due to the periodic domain structure. Diffracted intensity in the $HK$ plane around the $hk0$ (top panel) and $hk1$ (bottom panel) superlattice peaks for the case of (b) Ising and (c) artificially induced Bloch domain walls. The Bloch component gives rise an anisotropic diffuse scattering pattern around the $hk0$ peaks, with minima parallel to the in-plane component of the scattering vector, $Q_{IP}$. The flux-closure domain configuration gives rise to anisotropic diffuse scattering around the $hk1$ superlattice peaks with minima perpendicular to $Q_{IP}$. (d) and (e) show short range simulated diffraction patterns for the Ising and Bloch cases respectively, when both acoustic distortions and polar distortions are considered (top row), when only the effect of the acoustic distortion is isolated (middle row), and when only polar distortions are isolated (bottom row).

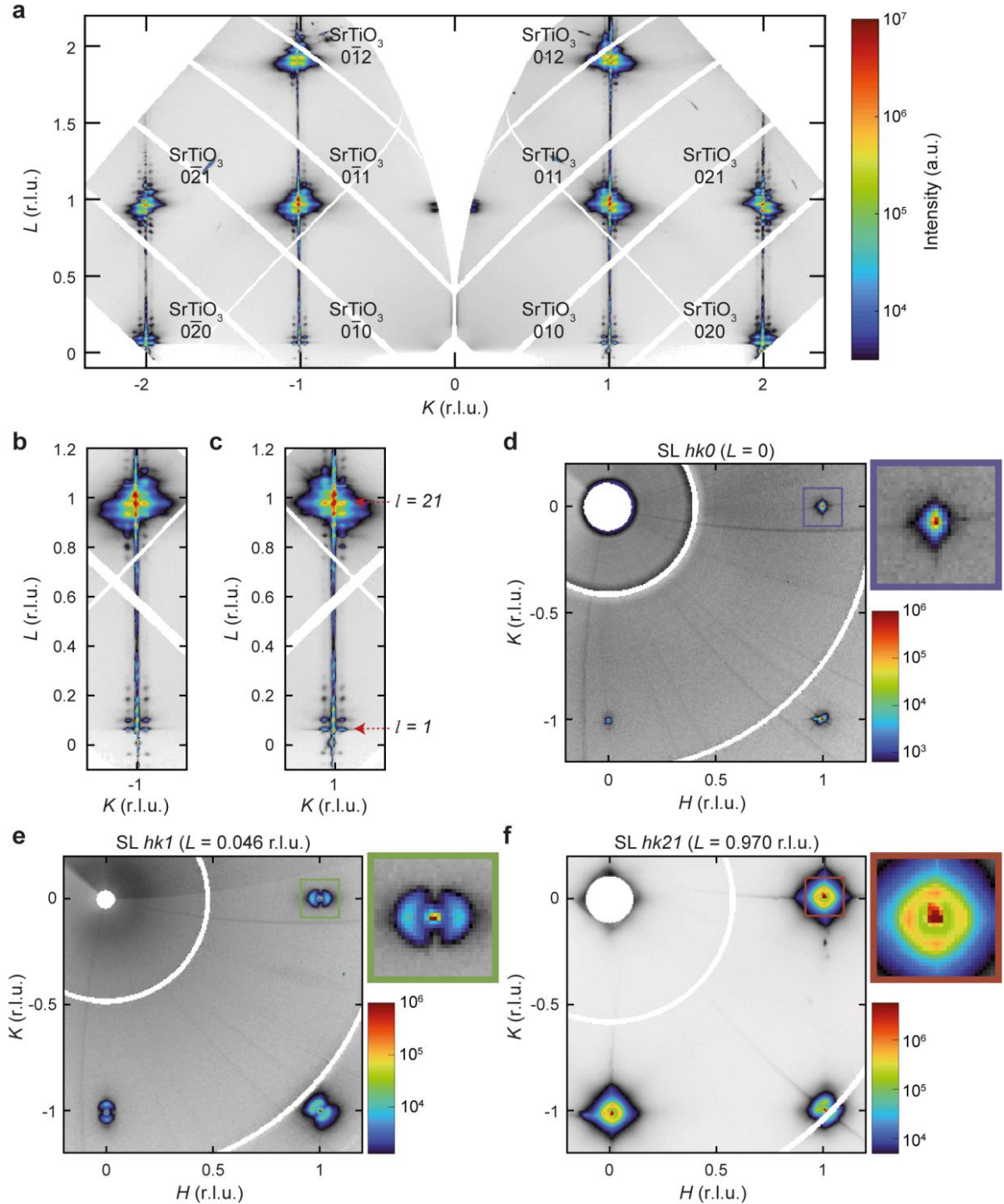

Figure 5. Diffuse scattering measurements performed at beamline ID28, ESRF. (a) Reciprocal space map in the $0KL$ plane of a $(PTO_{16}|STO_5)_{11}$ superlattice deposited on $SrTiO_3$. Shorter range maps around (b) $K = -1$ and (c) $K = 1$ show the $l = 1\ldots21$ superlattice peaks between $L = 0$ and $L = 1$. Diffuse signal appears around each superlattice Bragg peak, exhibiting a left-right asymmetry. The colour scale is identical to panel (a). (d) The $HK$ plane around the $hk0$ superlattice peaks, where no diffuse signal is observed. (e) Diffuse signal around the $hk1$ superlattice peaks, showing an anisotropic diffuse scattering pattern with minima along a direction perpendicular to $Q_{IP}$. (f) Diffuse signal around the $hk21$ superlattice peaks, showing a nearly isotropic diffuse scattering ring. The boxes in panels (d), (e) and (f) show high magnification views of the diffracted intensity around the 100, 101, and 1021 superlattice peaks. Note that the $1\bar{1}21$ peak is partially obscured by the detector gap. The reciprocal space maps are plotted using a hybrid colour scale to emphasise diffuse signal between Bragg peaks; the colour bar next to each panel is logarithmic and the intensity below the minimum value of each colour bar is expressed in linear grayscale. SL: superlattice. a.u.: arbitrary units.

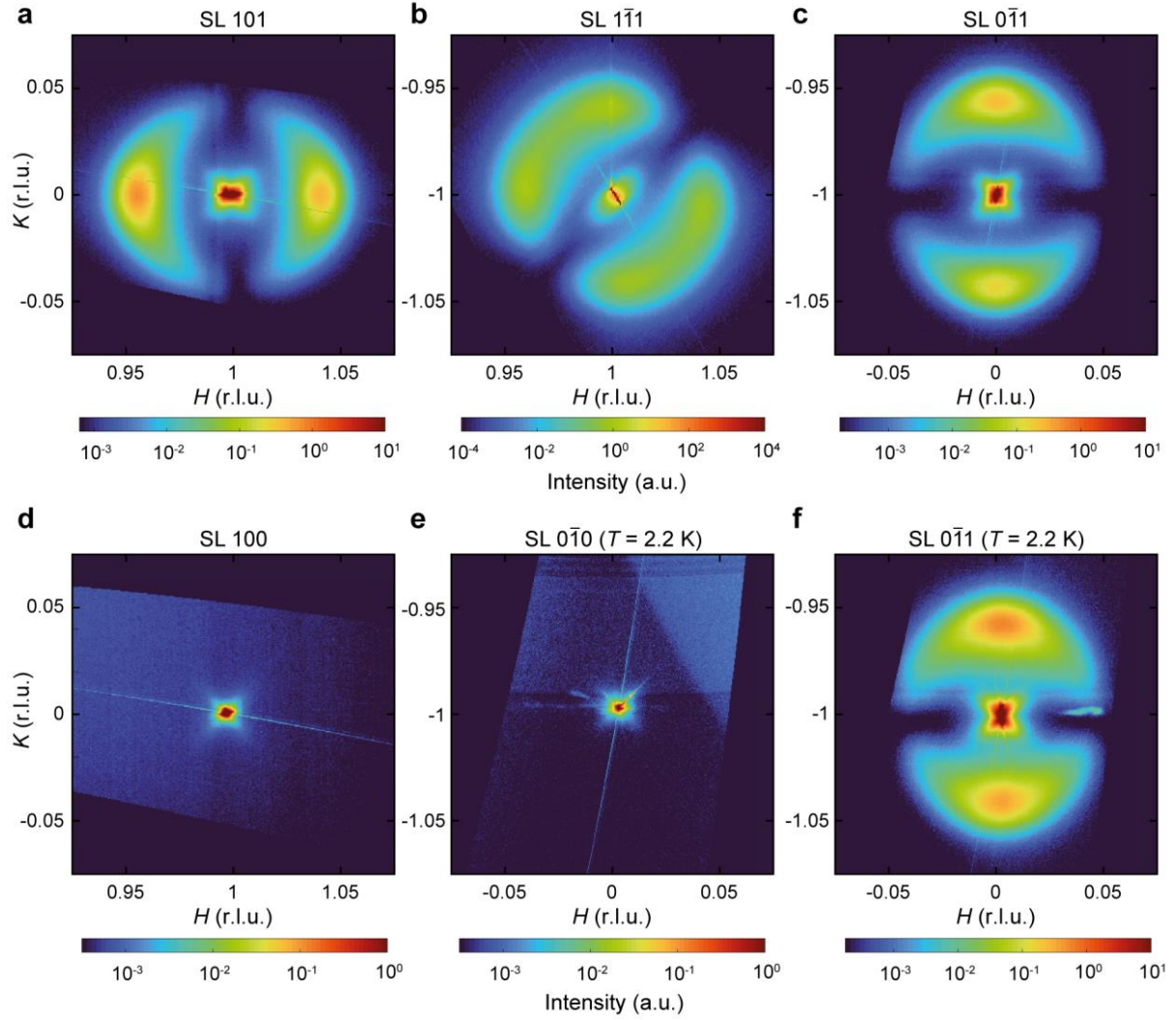

Figure 6. High-resolution measurements at beamline ID01, ESRF for a $(PTO_{13}|STO_5)_{12}$ superlattice. Reciprocal space maps in the $HK$ plane around the (a) 101 ($l = 1, L = 0.054$ r.l.u.), (b) $1\bar{1}1$, (c) $0\bar{1}1$, and (d) 100 ($l = 0, L = 0$) Bragg peaks of the superlattice, taken at room temperature. The minima in the diffuse scattering in panels (a) to (c) lie along a direction perpendicular to the in-plane component of the scattering vector, $Q_{IP}$. No diffuse scattering is observed around the 100 peak. (e) Reciprocal space map around the $0\bar{1}0$ peak at 2.2 K. No diffuse scattering is observed around this peak at low temperatures. (f) Reciprocal space map around the $0\bar{1}1$ peak at 2.2 K, showing no qualitative change in the diffuse scattering at low temperatures (quantitative changes in the diffuse scattering are summarised in the Supplementary Materials).

*Methods*

*First- and second-principles calculations*

Our DFT calculations are performed relying on the PBESol [49,50] exchange correlation functional as implemented in the ABINIT software [51]. The plane-wave energy cutoff was 65 Ha and the k-point grid was $8 \times 8 \times 8$ for the training set used to fit the second-principles model and was $8 \times 8 \times 1$ for the relaxation of the 60-atom cell. All structures were fully relaxed until the forces on the atoms were less than $10^{-5}$ Ha/Bohr and the stresses were less than $10^{-7}$ Ha/Bohr$^3$.

Our second-principles atomistic models of PbTiO$_3$ and SrTiO$_3$ have been constructed within the MULTIBINIT package and fitted on DFT data. The molecular dynamics (MD) calculations have been performed by using a hybrid Monte Carlo simulation within a $20 \times 12 \times 12$ supercell of the 5 atom unit cell of PbTiO$_3$. For each temperature, we performed 200,000 MD steps. Between each MD step, 40 Monte Carlo steps were performed in order to speed up the convergence and the energy landscape sampling. The local polarization has been calculated with the Born effective charge by averaging the displacements within the unit cell centred on the Pb atom following the procedure in Ref [20].

Our second-principles model for PbTiO$_3$/SrTiO$_3$ superlattice has been constructed following the scheme proposed in Ref. [28]. The reference lattice parameters for PbTiO$_3$/SrTiO$_3$ superlattices are calculated by minimising the elastic energy of the SrTiO$_3$ and PbTiO$_3$:

$$E = \frac{C_{STO}}{2}\left(\frac{a_{avg} - a_{STO}}{a_{STO}}\right)^2 + \frac{C_{PTO}}{2}\left(\frac{a_{avg} - a_{PTO}}{a_{PTO}}\right)^2$$

Where $C_{STO}(C_{PTO})$ are the harmonic elastic constants of SrTiO$_3$ (PbTiO$_3$) and $a_{STO}$ ($a_{PTO}$) are the lattice parameters of cubic SrTiO$_3$ (PbTiO$_3$). After minimization, the $a_{avg}$ can be calculated as:

$$a_{avg} = a_{STO}a_{PTO}\left(\frac{ma_{STO} + a_{PTO}}{ma_{STO}^2 + a_{PTO}^2}\right), \text{with } m = \frac{C_{STO}}{C_{PTO}}.$$

All structures were relaxed in order to minimize the force to $2 \times 10^{-5}$ Ha/Bohr and the stresses which were not constrained by expitaxial strain to be less than $2 \times 10^{-7}$ Ha/Bohr$^3$. In order to find the global minima of the system, we decrease the temperature from 10 K to 1 K in steps of 1 K. Between each temperature we perform 2000 MD steps using hybrid Monte Carlo: between each MD step, 40 Monte Carlo steps were performed.

*Phase-field simulations*

Phase-field simulations of PbTiO$_3$ based superlattices were carried out using the program Ferrodo [52,53], which allows to find stationary domain structure configurations in perovskite ferroelectrics defined by the generalized Ginzburg–Landau–Devonshire model [54]. In the spirit of the established Ginzburg–Landau–Devonshire approach to ferroelectric perovskites, the primary order-parameter field is the electric polarization $\boldsymbol{P(R)}$, which is subject both to anharmonic local interactions and long-range dipole-dipole interactions screened by an isotropic background permittivity. Furthermore, the Ginzburg–Landau–Devonshire energy functional contains terms linear or quadratic in the strain degrees of freedom. These secondary order parameters can be represented by the spatially inhomogeneous strain tensor field or the field of acoustic displacements $\boldsymbol{u(R)}$. Using the mechanical equilibrium conditions, the energy contributions containing strain degrees of freedom can be formally expressed as function of the polarization field $\boldsymbol{P(R)}$ with the average strain as a parameter. This formal solution is used to eliminate the secondary order parameters from the model so that the renormalized Ginzburg–Landau free-energy functional with the accordingly corrected coefficients and with the additional long-range interaction term depends on the primary order-parameter only. The renormalization is implicit in the tabulated Ginzburg-Landau model parameters so that technically, the renormalization consists in adding the long-range fields only. The temporal evolution $P(R)$ towards energy optimum is searched numerically by solving the time-dependent Ginzburg–Landau equation for such renormalized Ginzburg–Landau free energy.

The phase-field simulations were performed on a simulation box of $216 \times 216 \times 216$ sites consisting of twelve superlattice periods along the $z$ axis, each with 13 sites of PbTiO$_3$ and 5 sites of SrTiO$_3$. The spatial step was 0.4 nm. The domain structure in the main text was relaxed following 300,000 iterations from a white noise initial condition under periodic boundary conditions and epitaxial clamping conditions imposing the average in-plane strain ($e_{xx} = -0.015, e_{yy} = -0.015, e_{xy} = 0$), with the former values estimated from the mismatch between the experimental lattice parameter of SrTiO$_3$ and hypothetic cubic PbTiO$_3$ crystal at ambient conditions [55]. Ferroelectric PbTiO$_3$ was described by the same model as in ref. [56] applied to study PbTiO$_3$-paralectric superlattices. The Landau parameters at 298 K are: $\alpha_1 = -1.709 \times 10^8$ JmC$^{-2}$, $\alpha_{11} = -7.25 \times 10^7$ Jm$^5$C$^{-4}$, $\alpha_{12} = 7.5 \times 10^8$ Jm$^5$C$^{-4}$, $\alpha_{111} = 2.61 \times 10^8$ Jm$^9$C$^{-6}$, $\alpha_{112} = 6.1 \times 10^8$ Jm$^9$C$^{-6}$, $\alpha_{123} = -3.66 \times 10^9$ Jm$^9$C$^{-6}$, with gradient parameters $G_{11} = 1 \times 10^{-10}$ Jm$^3$C$^{-2}$, $G_{12} = -1 \times 10^{-10}$ Jm$^3$C$^{-2}$, $G_{44} = 1 \times 10^{-10}$ Jm$^3$C$^{-2}$, elastic components $C_{11} = 1.746 \times 10^{11}$ Jm$^{-3}$, $C_{12} = 7.94 \times 10^{10}$ Jm$^{-3}$, $C_{44} = 1.111 \times 10^{11}$ Jm$^{-3}$ and electrostriction parameters $q_{11} = 1.1412 \times 10^{10}$ JmC$^{-2}$, $q_{12} = 4.6 \times 10^8$ JmC$^{-2}$, $q_{44} = 7.5 \times 10^9$ JmC$^{-2}$, $Q_{11} = 0.089$ m$^4$C$^{-2}$, $Q_{12} = -0.026$ m$^4$C$^{-2}$, $Q_{44} = 0.0675$ m$^4$C$^{-2}$. The background permittivity $\varepsilon_B$, as defined in ref. [54,56], was set to 1. The Landau parameters of paraelectric SrTiO$_3$ were set as in ref. [57] ($\alpha_1 = 1.829 \times 10^8$ JmC$^{-2}$, $\alpha_{11} = 1.696 \times 10^9$ Jm$^5$C$^{-4}$, $\alpha_{12} = 3.918 \times 10^8$ Jm$^5$C$^{-4}$ at 298 K), and the gradient, elastic, electrostriction, and electrostatic terms were kept the same as its ferroelectric counterpart. The values of the dimensionless soft mode eigenvector ($d_{\mathrm{Pb}} = 0.0194$; $d_{\mathrm{Ti}} = -0.0163$; $d_{\mathrm{O1}} = d_{\mathrm{O2}} = d_{\mathrm{O3}} = -0.0802$) and of the effective flexoelectric tensor ($f_{11} = -6.1$ V, $f_{12} = -2.8$ V, $f_{44} = 2f_{1212} = -3.9$ V) have been determined using the second principles model for PbTiO$_3$ described in this work. The values were adjusted to best match with the corresponding correlations between the atomic displacements and polarization within a representative auxiliary multidomain configuration generated by incomplete annealing of a $4 \times 12 \times 20$ PbTiO$_3$ supercell of initially disordered displacements.

*Diffuse scattering calculations*

To calculate the diffuse scattering maps, the crystal structure was generated from the relaxed polarization field $P(R)$ by considering two contributions: local polar distortions $p_a(R)$ and local acoustic displacements $u(R)$. Local polar distortions $p_a(R)$ were expected to be directly proportional to the polarization field $P(R)$. The conversion of the local polarization $P(R)$ at the lattice vector $R$ to the atomic displacements within the corresponding perovskite unit cell was achieved using a linear relationship: $p_a(R) = d_a P(R) \left(\frac{a_0}{|P_s|}\right)$. Here, $a_0 = 0.4$ nm is a reference lattice constant, $P_s$ = 0.755 C/m$^2$ is the spontaneous polarization of monodomain PbTiO$_3$ at 298 K (as it appears in phenomenological Landau models), the index $a$ runs through the 5 atomic sublattices of the ABO$_3$ perovskite structure, and $d_a$ form dimensionless components of the soft mode eigenvector attached to the polarization in the Landau potential. The acoustic displacements $u(R)$ were calculated from the polarization field assuming mechanical equilibrium conditions with the help of the elastostatic Green function method [58] considering electrostriction and flexoelectric coupling to the polarization. The strain components $\varepsilon_{ij}$ were evaluated from displacements $u$ via $\varepsilon_{ij} = \frac{1}{2}\left(\frac{du_i}{dx_j} + \frac{du_j}{dx_i}\right)$, where $i,j = 1,2,3$. Both contributions to atomic displacements were used to generate an A-centered perovskite structure with the reference lattice constant 0.4 nm, which was used as an input to the *MP_tools* software package [59] to calculate x-ray scattering maps. Diffuse scattering data were symmetrized by considering the overall *4mm* symmetry.

*Sample growth*

PbTiO$_3$/SrTiO$_3$ superlattices were deposited on TiO$_2$-terminated (001) SrTiO$_3$ substrates (Crystec GmbH) using off-axis radiofrequency (RF) magnetron sputtering. The PbTiO$_3$ and SrTiO$_3$ layers were deposited from Pb$_{1.1}$TiO$_3$ and SrTiO$_3$ ceramic targets respectively, in a 0.18 Torr atmosphere with an oxygen to argon flow ratio of 20.4/28.7 and RF power of 60 W. During growth, the substrate was kept at a constant temperature of 560 °C, as measured by a thermocouple inside the heating block.

*Diffuse x-ray scattering measurements*

Diffuse x-ray scattering measurements were performed at the ID28 beamline (ESRF, France) in low incidence reflection geometry with incident photon wavelength $\lambda = 0.98$ Å [60]. All measurements were performed with the film rotated 45° perpendicular to the beam, minimising the impact of the excluded cone. Data were treated using the CrysAlis Pro software package (Rigaku Oxford Diffraction) and high-resolution reciprocal space reconstructions were produced using in-house beamline software. No symmetry averaging or space filling routines were applied. All measurements were conducted at room temperature.

*High-resolution synchrotron x-ray diffraction measurements*

High-resolution synchrotron x-ray diffraction measurements were performed at beamline ID01 at ESRF [61], using an incoherent X-ray beam (approximately 50 µm × 50 µm full width at half maximum) to probe the average domain behaviour. An incident X-ray energy of 9 keV was selected using a Si (111) double crystal monochromator. High-temperature measurements were performed with the sample mounted on a resistive heater using a high-temperature ceramic adhesive. Low-temperature measurements were performed by mounting the samples on a custom continuous flow liquid He cryostat. The samples were mounted on an aluminium stage using GE varnish. The temperature of the samples was monitored using a Cernox thermometer and controlled using a cartridge resistor (50 Ω, 25 W), both connected to a Lakeshore 340 temperature controller.

*Acknowledgments*

The authors thank Evgenios Stylianidis for help with sample fabrication, Pavel Márton for help with treating second principles data, and Jiří Kulda for his generous support with the *MP_tools* software package. This work was supported by the Swiss National Science Foundation (SNSF) Scientific Exchanges Scheme [Grant No. IZSEZ0_212990 (M.H.)], by Division II of the SNSF [Project No. 200021_178782 and 200021_178782 (C.L., L.D., J.-M.T., M.H.)], and the European Union's Horizon 2020 research and innovation program [Grant Agreement No. 766726 – TSAR (P.O., L.B., A.S., P.Z., P.G., J.H.)]. P.O. and J.H. acknowledge the assistance provided by the Ferroic Multifunctionalities project, supported by the Ministry of Education, Youth, and Sports of the Czech Republic. Project No. CZ.02.01.01/00/22_008/0004591, co-funded by the European Union. P.G. acknowledges support from F.R.S.-FNRS Belgium (Grant No T.0107.20, PROMOSPAN). We acknowledge the European Synchrotron Radiation Facility and the ID01 and ID28 beamline staff for support during the synchrotron experiments. For simulations, we acknowledge access to the CECI supercomputer facilities funded by the F.R.S.-FNRS Belgium (Grant No. 2.5020.1) and to the Tier-1 supercomputer of the Fédération Wallonie-Bruxelles funded by the Walloon Region of Belgium (Grant No. 1117545).

**SUPPLEMENTARY INFORMATION**

Edoardo Zatterin[1†], Petr Ondrejkovic[2†], Louis Bastogne[3†], Céline Lichtensteiger[4], Ludovica Tovaglieri[4], Daniel A. Chaney[1], Alireza Sasani[3], Tobias Schülli[1], Alexei Bosak[1], Steven Leake[1], Pavlo Zubko[5,6], Philippe Ghosez[3], Jirka Hlinka[2], Jean-Marc Triscone[4], Marios Hadjimichael[4,7*]

1 ESRF—The European Synchrotron, 71 Avenue des Martyrs, 38000 Grenoble, France
2 Institute of Physics of the Czech Academy of Sciences, Na Slovance 2, 18221 Praha 8, Czech Republic
3 Theoretical Materials Physics, Université de Liège, Allée du 6 août, 17, B-4000 Sart Tilman, Belgium
4 Department of Quantum Matter Physics, University of Geneva, 24 Quai Ernest-Ansermet, 1211 Geneva, Switzerland
5 Department of Physics and Astronomy, University College London, Gower Street, London WC1E 6BT, United Kingdom
6 London Centre for Nanotechnology, 17-19 Gordon Street, London WC1H 0AH, United Kingdom
7 Department of Physics, University of Warwick, Coventry, CV4 7AL, United Kingdom
*Corresponding author: marios.hadjimichael@warwick.ac.uk
†These authors contributed equally: Edoardo Zatterin, Petr Ondrejkovic, Louis Bastogne

*Supplementary Part 1: Equilibrium domain period from second-principles calculations*

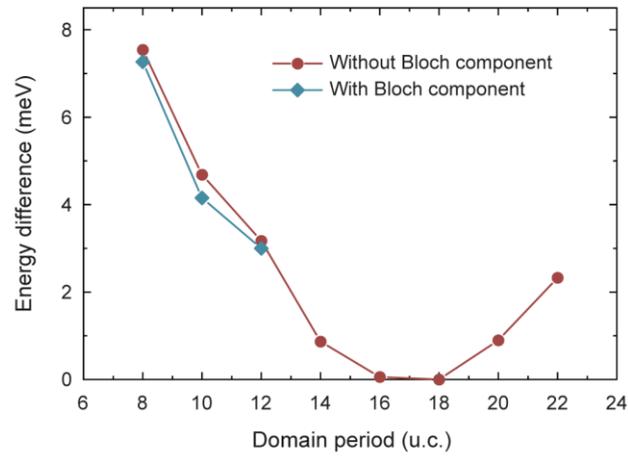

Figure S1. Equilibrium domain period for the (PTO$_6$|STO$_6$) superlattice extracted using second-principles calculations. Plot of the energy difference with respect to the minimum energy domain period (18 unit cells) as a function of domain period. Red circles and blue diamonds are data points for domain walls without a Bloch polarisation and with a Bloch polarisation respectively.

*Supplementary Part 2: Expected diffraction pattern schematics*

The periodic modulation of the atomic positions in PbTiO$_3$ due to the nanoscale ferroelectric domain structure leads to the appearance of additional diffuse intensity around the thin film Bragg peaks [1,2]. For signal around the $00L$ Bragg peaks, the intensity distribution of the diffuse scattering in the $HK$ plane is determined by the distribution of domain-wall orientations [3,4]. In the case of PbTiO$_3$ deposited on low-miscut SrTiO$_3$, the domains adopt an almost isotropic, labyrinth-like pattern, which can manifest as an isotropic ring of intensity around the superlattice Bragg peaks [1–3,5,6]. Some preference for (100) domain walls can occur in these systems [3], but for the purposes of this section we assume that the diffuse intensity is completely isotropic.

The expression for the diffracted amplitude in the main text (Equation 1) shows that diffuse scattering due to periodic domains can be observed when the scalar product between the scattering vector $\boldsymbol{Q}$ and the cation displacement $\delta \boldsymbol{r}_j$ is non-zero. When the out-of-plane component of the scattering vector, $Q_z$, is zero, the diffracted amplitude is sensitive to in-plane modulations of atomic positions. In the case of stripe domains with a domain wall polarisation in the plane of the wall (and thus an ionic displacement perpendicular to the direction of the domain periodicity), $\boldsymbol{Q} \cdot \boldsymbol{P}$ will be zero for modulation directions parallel to $\boldsymbol{Q}$. Conversely, for domain wall polarisation perpendicular to the plane of the wall, $\boldsymbol{Q} \cdot \boldsymbol{P}$ will be zero when the direction of the domain periodicity is perpendicular to $\boldsymbol{Q}$. Figures S1(a) and S1(b) show the corresponding intensity distributions expected for these two scenarios around the $100$, $0\bar{1}0$ and $1\bar{1}0$ peaks of a thin film or superlattice with a periodic domain structure. The schematics assume, for simplicity, that the domain structure is labyrinth-like with an isotropic distribution of domain wall orientations, and that the magnitude of the in-plane atomic displacements within the domain walls does not depend on the domain wall orientation.

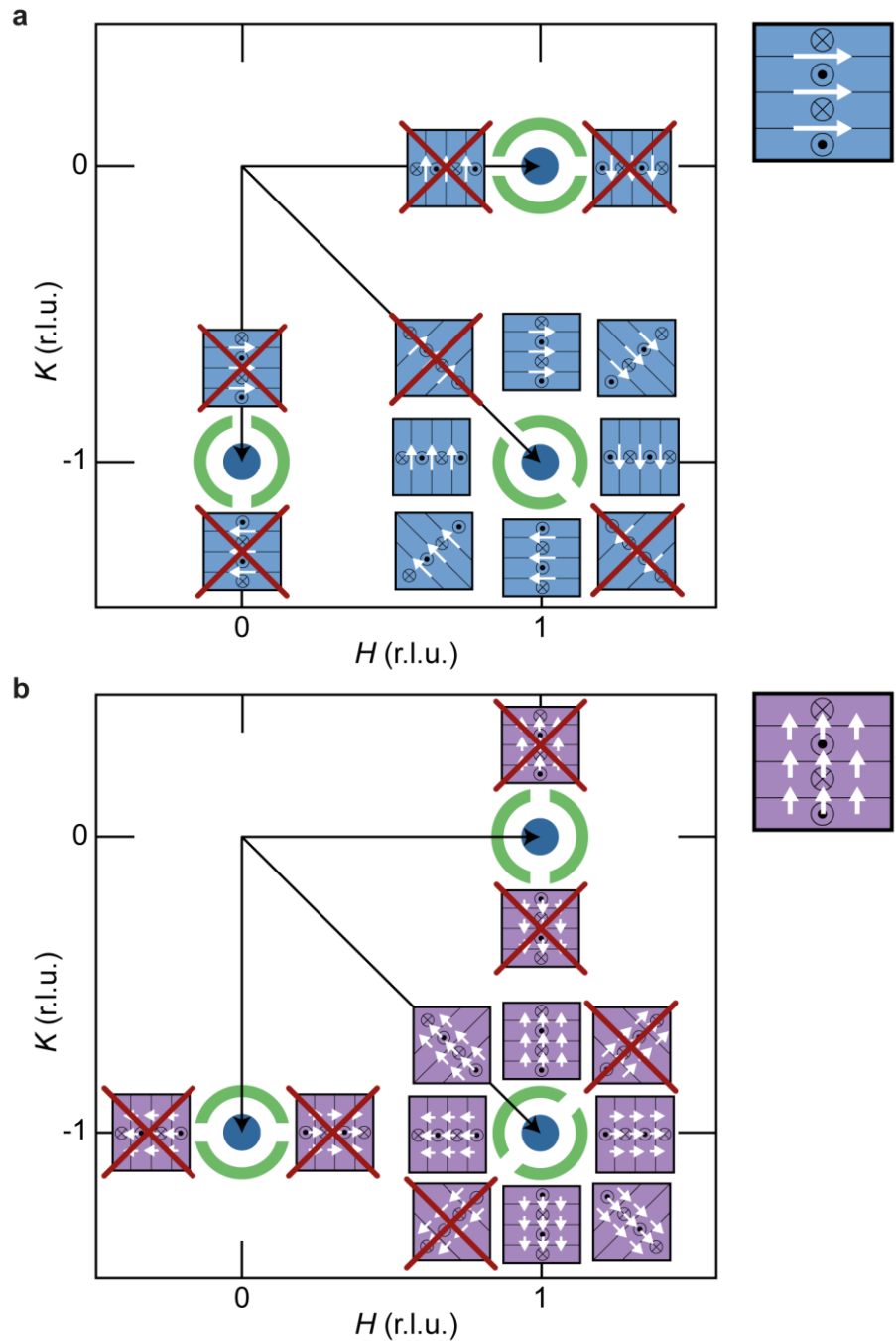

Figure S2. Expected diffraction patterns for cases in which the domain walls exhibit a polarisation parallel or perpendicular to their planes. (a) Expected diffraction pattern in the $HK0$ plane for a periodic ferroelectric domain structure with ordered polarisation parallel to the domain wall planes. The schematics show that we are not sensitive to modulations of the polarisation for periodicities along the in-plane component of the scattering vector, $Q_{IP}$, giving rise to minima in the diffuse scattering in this direction. (b) Expected diffraction pattern in the $HK0$ plane for a ferroelectric domain structure with ordered polarisation perpendicular to the domain wall planes. The minima in diffuse scattering now occur in a direction perpendicular to $Q_{IP}$.

*Supplementary Part 3: Diffuse scattering asymmetry*

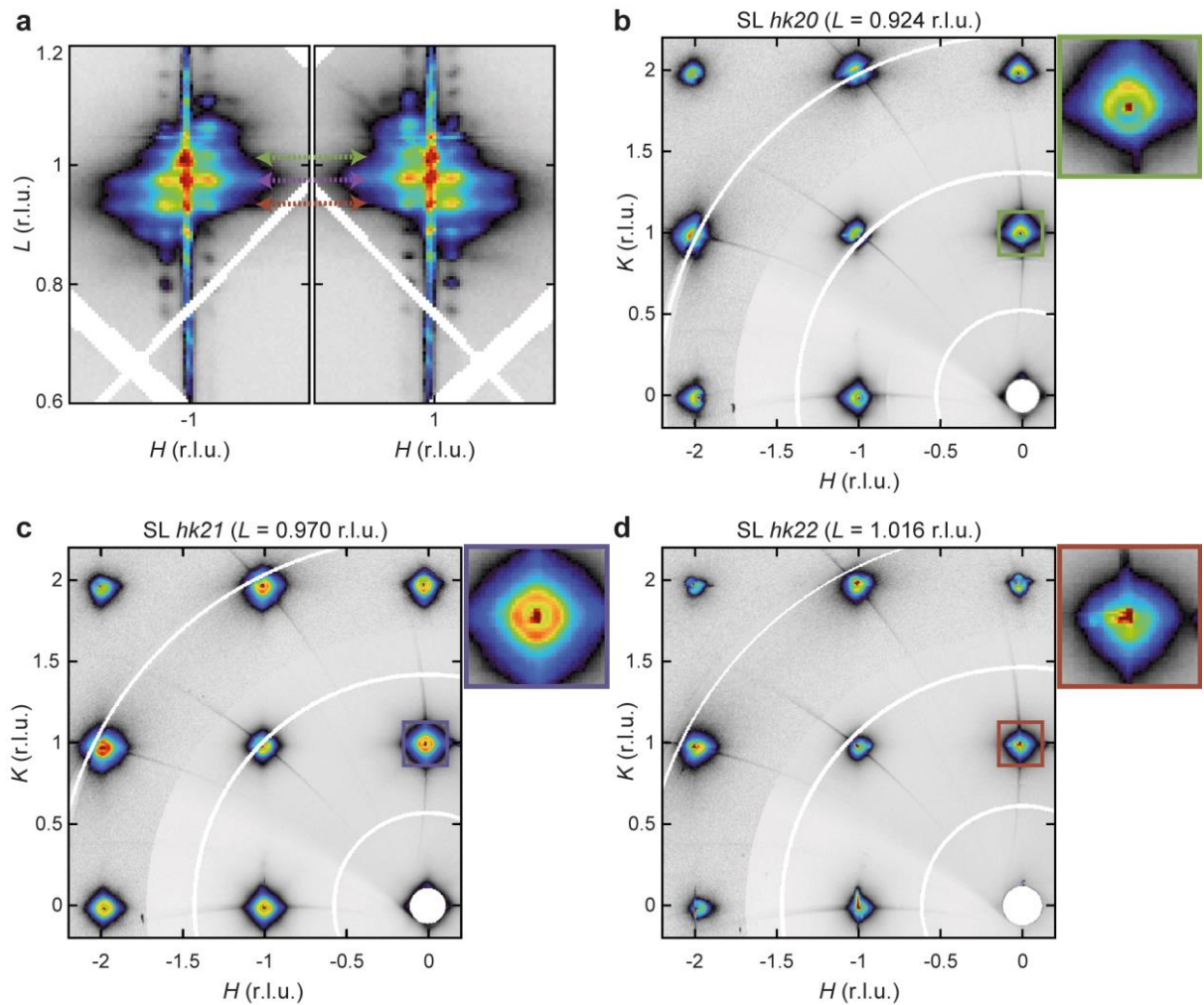

Figure S3. Asymmetry of the diffuse intensity near the $L = 1$ substrate peak. (a) Reciprocal space maps in the $H0L$ plane of a (PTO$_{16}$|STO$_5$)$_{11}$ superlattice deposited on SrTiO$_3$, showing the region around $H = \pm 1$ and $L = 1$. The horizontal dashed arrows show the different $L$ positions for the $hk20$, $hk21$ and $hk22$ superlattice peaks. (b) The $HK$ plane around the $hk20$ superlattice peaks, showing that the diffuse intensity further away from the origin ($H = K = 0$) is more intense than nearer the origin. (c) The $HK$ plane around the $hk21$ superlattice peaks, where the diffuse intensity becomes more isotropic. (d) The $HK$ plane around the $hk22$ superlattice peaks, where the diffuse intensity nearer the origin is more intense compared to that further from the origin.

*Supplementary Part 4: Comparison of 1 0 54 and 101 superlattice peaks*

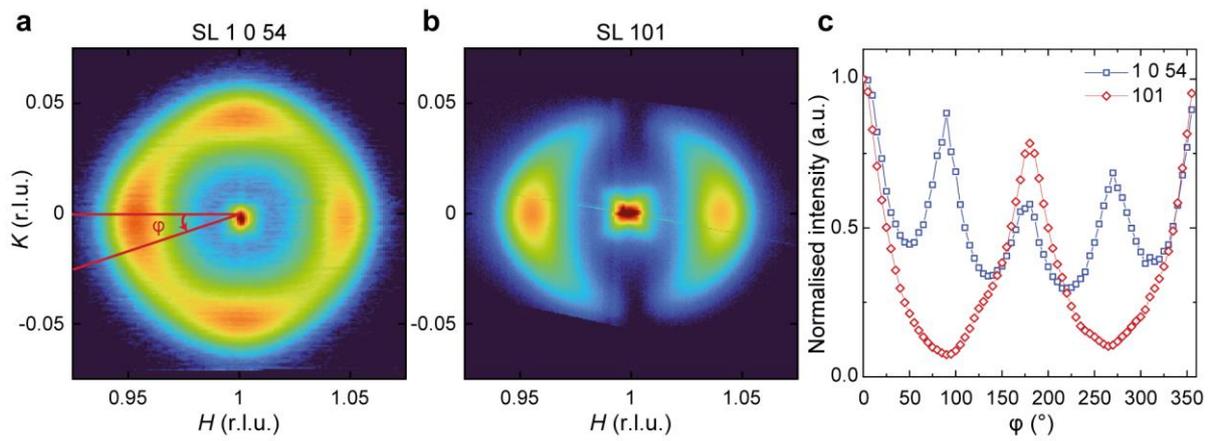

Figure S4. Comparison of 1 0 54 and 101 superlattice peaks. Reciprocal space maps in the $HK$ plane around the (a) 1 0 54 ($l = 54, L = 2.935$ r.l.u) and (b) 101 ($l = 1, L = 0.054$ r.l.u.) Bragg peaks of a $(PTO_{13}|STO_5)_{12}$ superlattice. (c) Plot of the diffuse scattering intensity as a function of azimuthal angle $\phi$, defined in panel (a). The plot shows that four maxima in intensity occur at $\phi = 0°, 90°, 180°, 270°$ in the diffuse scattering around the 1 0 54 peak. When the azimuthal distribution of the diffuse scattering around the 101 peak is plotted, we observe minima in intensity at $\phi = 90°$ and $\phi = 270°$.

*Supplementary Part 5: Disappearance of diffuse scattering at high temperatures*

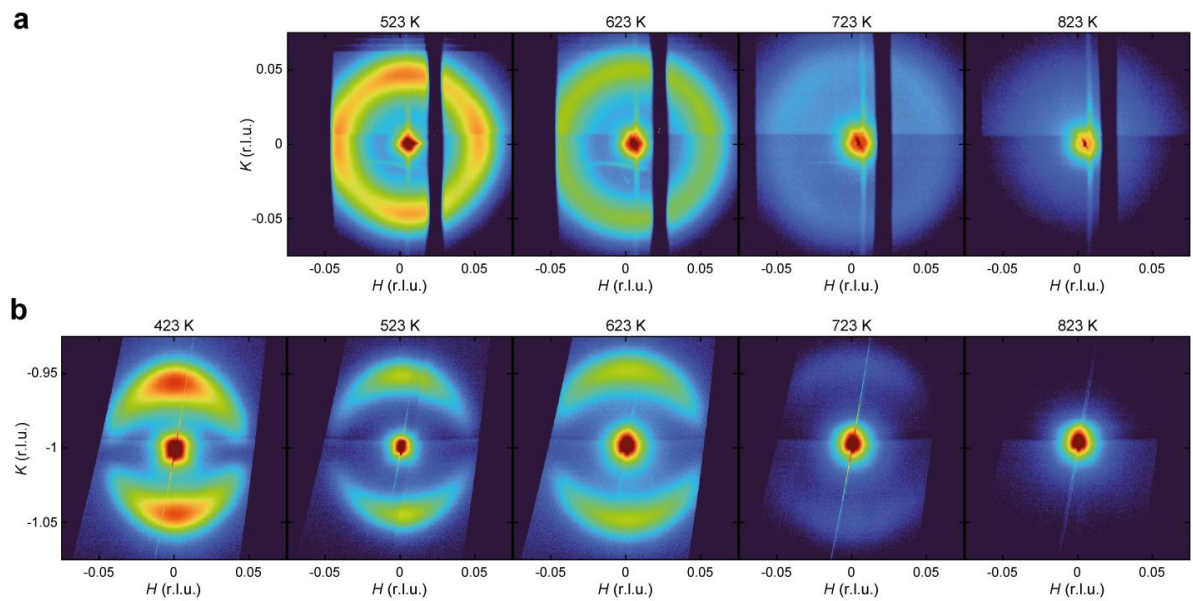

Figure S5. High-temperature behaviour of the diffuse scattering around the Bragg peaks of a $(PTO_{13}|STO_5)_{12}$ superlattice deposited on $SrTiO_3$. Reciprocal space maps in the HK plane around the (a) 0036 and (b) $0\bar{1}1$ superlattice peaks, showing the disappearance of the periodic satellites around both peaks above 723 K.

*Supplementary Part 6: Low-temperature x-ray diffraction measurements*

Figure S6 shows reciprocal space maps around the $0\bar{1}1$ superlattice peak for a $(PTO_{13}|STO_5)_{12}$ superlattice at temperatures from 2.2 K up to 125 K. We observe that there is no striking change in the diffuse signal as the temperature is decreased down to 2.2 K, and the diffuse signal remains consistent with a flux-closure configuration. Additionally, we note that no diffuse signal appears around the $0\bar{1}0$ superlattice peak down to 2.2 K (Figure S6(d)).

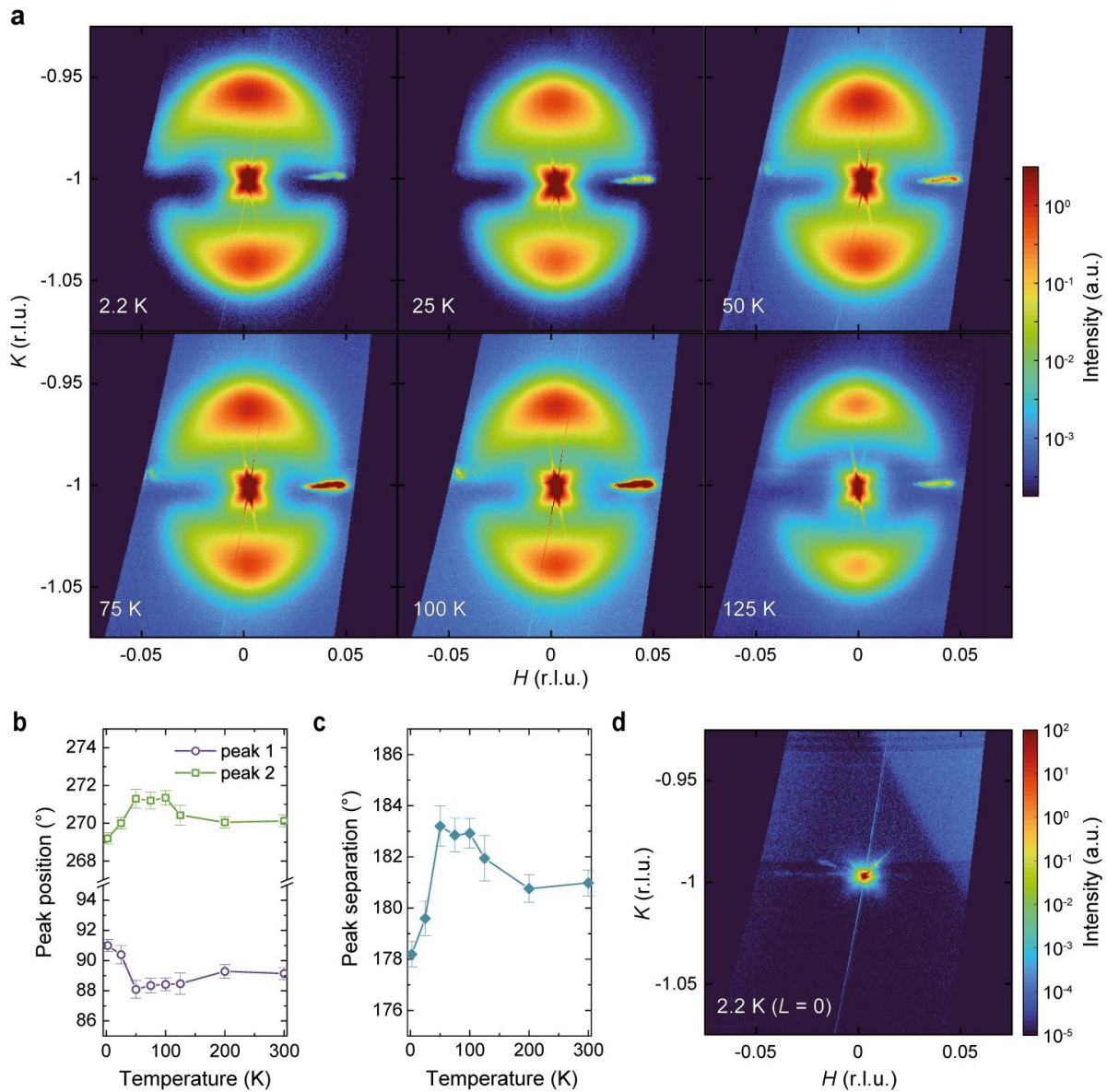

Figure S6. Low-temperature measurements. (a) Reciprocal space maps in the $HK$ plane as a function of temperature around the $0\bar{1}1$ ($l=1, L=0.054$ r.l.u.) Bragg peak of a $(PTO_{13}|STO_5)_{12}$ superlattice. No qualitative difference is observed in the diffuse scattering patterns down to 2.2 K. (b) Azimuthal position of main diffuse scattering peaks around the $0\bar{1}1$ Bragg peak of the superlattice as a function of temperature. (c) Angular separation between the two diffuse scattering peaks as a function of temperature. (d) Reciprocal space map in the $HK$ plane around the $0\bar{1}0$ peak at 2.2 K showing no appearance of diffuse scattering at low temperatures.

Interestingly, a closer examination of the diffuse scattering around the $0\bar{1}1$ peak of the superlattice at 2.2 K shows that the diffuse scattering intensity is no longer symmetric around $H = 0$, as the satellite maxima are no longer aligned with the [010] direction. In an attempt to understand this subtle change further, we plot the intensity of the diffuse scattering as a function of azimuthal angle ϕ and fit the position of each of the two peaks using a pseudo-Voigt function (the angular positions of the two peaks are plotted in Figure S6(b)). Figure S6(b) shows that the position of the first peak starts shifting towards higher angles at approximately 50 K, while at that temperature the position of the second peak starts shifting towards lower angles (the separation between the two peaks therefore starts decreasing at this temperature as shown in Figure S6(c)). Similarly, this asymmetry occurs for superlattices with other PbTiO$_3$ thicknesses, with larger degree of asymmetry for the superlattice with thicker PbTiO$_3$ layers (as shown in Figure S7).

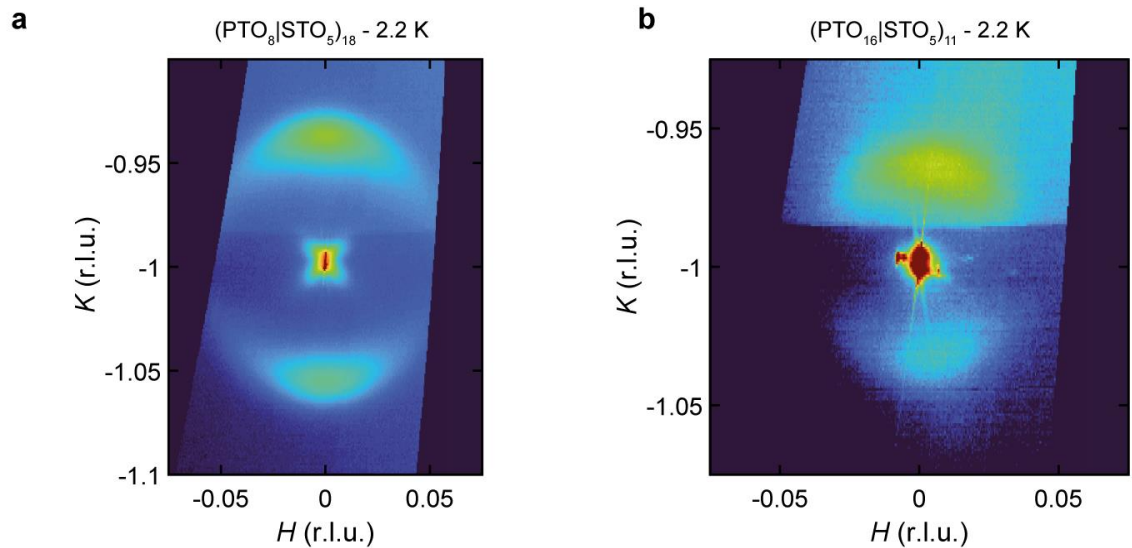

Figure S7. Low-temperature measurements for superlattices with different PbTiO$_3$ thicknesses. (a) Reciprocal space maps in the $HK$ plane around the $0\bar{1}1$ Bragg peaks of a (a) (PTO$_8$|STO$_5$)$_{18}$ and a (b) (PTO$_{16}$|STO$_5$)$_{11}$ superlattice, with domain periods equal to 67 and 110 Å respectively, taken at 2.2 K. The asymmetry in the diffuse scattering peaks is more pronounced in the superlattice with thicker PbTiO$_3$ layers.

A similar type of asymmetry in the periodic satellites was observed in surface-sensitive RSXD studies of magnetic multilayers, in a region near the sample surface where an uncompensated Néel and Bloch magnetisation configuration coexisted at a domain wall [7]. However, this asymmetry was not explored further in this work and the reason for its appearance was not determined. The change in peak separation in our measurements could be related to a slight change in the structure of the domain walls, or possibly a change in the structure of PbTiO$_3$ at low temperatures [8–10]. It could also be related to the low-temperature behaviour of the SrTiO$_3$ substrate, which undergoes a cubic to tetragonal transition at approximately 110 K [11] and could affect the ferroelectric domain structure. As the reason for this change in the diffuse scattering is still unknown, the low-temperature behaviour of the 180° domain walls in PbTiO$_3$ deserves further investigation.